\documentclass[preprint]{aastex}

\shorttitle{$FUSE$ Observations of 3C\,273}
\shortauthors{Sembach et al.}

\begin{document}

\newcommand{\kms}{\,km\,s$^{-1}$}     

\title{Far Ultraviolet Spectroscopy of the
Intergalactic and Interstellar Absorption Toward 3C\,273}

\slugcomment{To appear in the Nov. 10, 2001 issue of {\it The Astrophysical Journal}}

\author{Kenneth~R.~Sembach\altaffilmark{1}, 
        J.~Christopher~Howk\altaffilmark{1}, 
        Blair~D.~Savage\altaffilmark{2}, 
	J.~Michael~Shull\altaffilmark{3},
	and William~R.~Oegerle\altaffilmark{4}} 
\altaffiltext{1}{Department of Physics and Astronomy, The Johns 
	Hopkins University, Baltimore, MD  21218}
\altaffiltext{2}{Department of Astronomy, University of Wisconsin, Madison, 
	WI  53706}
\altaffiltext{3}{CASA and JILA, Department of Astrophysical and Planetary Sciences,
University of Colorado, Boulder, CO  80309}
\altaffiltext{4}{Laboratory for Astronomy and Solar Physics, Code 681, Goddard Space Flight Center, Greenbelt, MD 20771}

\begin{abstract}
We present {\it Far Ultraviolet Spectroscopic Explorer} observations of the 
 molecular, neutral atomic,
weakly ionized, and highly ionized components of the interstellar and 
intergalactic material toward the quasar 3C\,273.
We identify Ly$\beta$ absorption in 8 of the known intergalactic
Ly$\alpha$ absorbers along the sight line with rest-frame equivalent
widths $W_{r}{\rm (Ly}\alpha)
 \gtrsim 50$ m\AA.  Refined estimates of the 
\ion{H}{1} column densities and Doppler parameters ($b$) of the clouds are 
presented.  We find a range of $ b  \approx 16-46$ \kms.
We detect multiple \ion{H}{1} lines (Ly$\beta$ -- Ly$\theta$) 
in the 1590 \kms\ Virgo absorber and estimate 
$\log N$(\ion{H}{1}) =
$15.85\pm^{0.10}_{0.08}$, ten times more \ion{H}{1} than all of 
the other absorbers along the sight line 
combined.  The Doppler width of this absorber, $b \approx 16$ \kms,
implies $T \lesssim 15,000$\,K.
We detect \ion{O}{6} absorption at 1015 \kms\ at the $2-3\sigma$ level that
may be associated with hot, X-ray emitting gas in the Virgo Cluster.
We detect weak \ion{C}{3}  and \ion{O}{6} absorption in the IGM at $z=0.12007$;
this absorber is predominantly ionized and has 
N(H$^+$)/N(\ion{H}{1}) $\ge 4000\,Z^{-1}$, where $Z$ is the metallicity. 

Strong Galactic interstellar \ion{O}{6} is present
between $-100$ and $+100$ \kms, with an additional 
high-velocity wing containing about 13\% of the total \ion{O}{6} between 
$\sim+100$ and $\sim+240$ \kms.  
The Galactic \ion{O}{6}, \ion{N}{5}, and \ion{C}{4} lines have similar
shapes, with roughly constant ratios across the --100 to +100 \kms\
velocity range.  The high velocity \ion{O}{6} wing is not detected in other 
species.  Much of the interstellar high ion absorption probably
occurs within a highly fragmented medium within the Loop~IV remnant or in the 
outer cavity walls of the remnant. Multiple hot gas production mechanisms 
are required.  The broad \ion{O}{6}
absorption wing likely traces the expulsion of hot gas out of the Galactic 
disk into the halo.
A flux limit of $5.4\times10^{-16}$ erg\,cm$^{-2}$\,s$^{-1}$
on the amount of diffuse \ion{O}{6} emission present $\sim3.5$\arcmin\
off the 3C\,273 sight line
combined with the observed \ion{O}{6} column density toward 3C\,273, 
log\,N(\ion{O}{6}) = $14.73\pm0.04$, implies 
$n_e \lesssim 0.02$ cm$^{-3}$ and P/k $\lesssim 11,500$ cm$^{-3}$~K for 
an assumed temperature of $3\times10^5$~K.
The elemental abundances in the neutral and weakly-ionized interstellar clouds 
are similar to those found for other halo clouds. The warm neutral and warm 
ionized clouds along the sight line have similar dust-phase abundances, 
implying that the properties of the dust grains in the two types of clouds 
are similar.
   Interstellar H$_2$ absorption is present at positive velocities at a level 
   of $\log N$(H$_2$) $\sim 15.71$, but is very weak at the velocities of the 
   main column density concentration along the sight line observed in 
   \ion{H}{1} 21\,cm emission.
\end{abstract}


\section{Introduction}
The 3C\,273 sight line ($l = 289.95\degr$, $b = +64.36\degr$, 
$z_{QSO} = 0.1583$)
is one of the best studied directions in the sky at ultraviolet wavelengths.
It has been the subject of numerous spectroscopic investigations 
encompassing studies of the ionization, kinematics, and chemical 
composition of the Milky Way interstellar medium (ISM; York et al. 1984; 
Burks et al. 1991, 1994; Savage et al. 1993;
Sembach, Savage, \& Tripp 1997), the content and 
evolution of the intergalactic medium (IGM; Morris et al. 1991; Bahcall 
et al. 1991, 1993; Brandt et al. 1993, 1997; Weymann et al. 1995; 
Hurwitz et al. 1998; Penton, Shull, \& Stocke 2000; Penton, Stocke, \& Shull
2000), and 
the relationship of the intergalactic Ly$\alpha$ absorption features to 
galaxies (e.g., Morris et al. 1993; Penton, Stocke, \& Shull 2001).  
The sight line contains 18 
intergalactic Ly$\alpha$ gas clouds at redshifts 
$z = 0.00338 - 0.14560$.  In this paper we present
new {\it Far Ultraviolet 
Spectroscopic Explorer} ($FUSE$) observations of 3C\,273 that expand upon
the previous investigations of the ISM and IGM in this direction.

$FUSE$ is the first far-ultraviolet ($\lambda < 1200$\,\AA) 
observatory to 
have both the sensitivity and spectral resolution necessary to perform 
detailed investigations of the kinematics and composition of the ISM and IGM 
in the directions of QSOs and active galactic nuclei (AGNs).  The $FUSE$
bandpass contains important diagnostics of molecular, atomic, and ionized
gases (see Sembach 1999).  Of paramount importance is the ability of $FUSE$ to 
detect the higher Lyman series lines 
of \ion{H}{1} and  
ionized gas lines of \ion{C}{3} $\lambda977.020$ and \ion{O}{6}
$\lambda\lambda1031.926, 1037.617$ in the Milky Way ISM and low-redshift
universe.  $FUSE$ investigations of intergalactic 
absorption and hot gas in the Galaxy
include studies of the low-redshift Ly$\beta$ forest (Shull et al. 2000a),
Milky Way halo gas (Savage et al. 2000; Shull et al. 2000b),  
high velocity clouds (HVCs; Sembach et al. 2000b, 2001; Richter et al. 2001b), 
and the individual sight lines toward H~1821+643 
(Oegerle et al. 2000), PG~0804+761 (Richter et al. 2001a), and 
PG~0953+415 (Savage et al. 2001a).

In this paper, we use $FUSE$ data to
investigate the IGM and ISM along the 3C\,273 sight 
line.  In \S2 we describe the observations and 
data reduction procedures.  Section 3 contains information
about absorption features in the $FUSE$ spectra, including 
equivalent width measurements, column density derivations, and models used to 
describe the absorption lines.
In \S4 we present results for the intergalactic absorption along the 
sight line, and in \S5 we describe the Milky Way absorption.  Section 6
contains a discussion of these results.  Section 7 consists of a brief summary
of the scientific results of this investigation.

\section{Observations}

\subsection{Far Ultraviolet Spectroscopic Explorer Observations}

$FUSE$ contains four Rowland circle spectrographs that produce
spectra over the wavelength range 905--1187\,\AA.  Two of the spectrographs 
have optics 
(a holographically-ruled diffraction grating illuminated by an off-axis 
paraboloid mirror) coated with 
Al+LiF for greatest sensitivity above $\sim1000$\,\AA, and two have optics 
coated with SiC for maximum sensitivity below $\sim1000$\,\AA.  The 
dispersed spectra from the four channels (LiF1, LiF2, SiC1, SiC2)
are recorded by two delay-line microchannel-plate detectors,
each individually covering nearly the full $FUSE$ bandpass.  Descriptions of 
the $FUSE$ instrumentation and its on-orbit performance are given by
Moos et al. (2000) and Sahnow et al. (2000a,b).

We obtained a 42.2 ksec observation of QSO 3C\,237 consisting of 30 
individual exposures on 2000~April~23, with the light of the quasar centered 
in the 
large (30\arcsec$\times$30\arcsec) apertures of the four optical channels.  
During the course of the observation, the light remained centered in the LiF1, 
LiF2, and SiC2 apertures but moved in and out of the SiC1 aperture due to 
thermally-induced image motions in the optical system.  As a result, the net 
exposure time in the SiC1 channel was only 12.6 ksec, and the resulting 
SiC1 spectral resolution was degraded slightly.  Most of the observation 
($\sim30$ ksec) took place during orbital night.

We processed the data using the standard CALFUSE (v1.8.7) pipeline software
available at the Johns Hopkins University in November 2000.  The software 
screened the raw photon address lists for valid events satisfying constraints 
imposed for Earth limb avoidance, South Atlantic Anomaly passage, and pulse
height distribution.  It also corrected for large scale geometric distortions 
in the detector grid, changes in spectral positions on the detector due to 
minute grating movements induced by thermal variations, Doppler shifts 
caused by orbital motion of the satellite, and detector 
background noise (which is 
minimal at the flux levels considered in this study).  The two-dimensional 
astigmatic spectra were extracted with a slit width appropriate for the 
spectral height at each wavelength.  There was no explicit correction
for the astigmatism available at the time of processing, which resulted in 
a slightly lower spectral resolution than could be achieved if the spectra 
were rectified before being summed in the spatial direction.  By 
spreading the spectral information over many detector pixels in the spatial 
direction, however, the astigmatism reduces the detector fixed-pattern noise 
in the extracted one-dimensional spectra.  No explicit 
flat-field corrections were applied in the standard $FUSE$ processing, 
so we made use of the overlapping
wavelength coverage of the different channels to determine whether 
fixed-pattern noise artifacts affect the lines used in the analysis described
in \S3 and \S4.

We determined the zero point of the wavelength scale by 
registering the molecular (H$_2$) and neutral and singly ionized atomic
lines (\ion{Si}{2} 
$\lambda1020.699$; \ion{Ar}{1} $\lambda1048.220,1066.660$; \ion{Fe}{2}
$\lambda1125.448,1143.226,1144.938$) in the spectrum to the 
longer wavelength \ion{S}{2}~$\lambda1250.584,1253.811$
lines observed with the Hubble Space Telescope ($HST$) and the \ion{H}{1}
21\,cm emission profile observed with the NRAO 140\arcmin\
telescope (Murphy, Sembach, \& Lockman, unpublished).  We found that 
in some wavelength regions (e.g., 
$\sim1040-1070$\,\AA) there is a slight offset of the interstellar lines from 
similar lines 
at other wavelengths.  This may be due to a slight stretch in the wavelength
dispersion solution introduced by alignment differences
between the wavelength calibration spectra and the 3C\,273 spectrum,
detector non-uniformities, or other unknown causes.  Whenever possible,
we have corrected for these relative wavelength shifts and believe that
the residual uncertainties are of the order of $\lesssim$ 8 \kms\ 
(1$\sigma$).  There is no evidence that these uncertainties are present on
wavelength scales comparable to the widths of the observed absorption lines 
(i.e., the line
profile shapes are not affected by uncertainties in the wavelengths).

The $FUSE$ data are oversampled at the current instrumental resolution.  
Therefore, we rebinned the fully-sampled ($\sim0.0065$ \AA\ pix$^{-1}$)
data by a factor of 3, which results in $\sim$3--4 bins per $\sim$17--22
\kms\ resolution element.
The data have $S/N$ 
$\approx 15-35$ per resolution element, depending upon wavelength and 
detector segment.  

The fully reduced data are shown in Figure~1, in which we plot the 
spectra from the detector segments having the highest signal-to-noise ratio.  
The coverage shown is as follows: SiC1A ($1079-1089$\,\AA), 
SiC1B ($\lambda \le 992$\,\AA), SiC2A ($\lambda \le 1001$\,\AA), 
SiC2B ($1075-1088$\,\AA), LiF1A ($994-1081$\,\AA), LiF1B ($1095-1186$\,\AA), 
LiF2A ($1088-1182$\,\AA), and LiF2B ($999-1073$\,\AA). 
In general, the SiC2 data 
are of better quality than the SiC1 data for this observation because of the 
alignment difficulties encountered for the SiC1 channel.  
Prominent interstellar and 
intergalactic absorption lines are indicated by tick marks above the spectra
and are identified to the right of each panel.  Many H$_2$ lines in rotational
levels $J = 0-3$ are indicated; some of the weaker lines in these levels 
overlap other stronger features and are not explicitly marked.  The expected
locations of weak intergalactic features that are not readily apparent in the 
spectra are identified with an additional ``?'' appended to the redshift.
At some wavelengths, 
fixed-pattern noise (FPN) features are present (e.g., 997.2\,\AA,
1060.7\,\AA\, 1070.9\,\AA, 1140.5\,\AA, and 1151.9\,\AA)
as evidenced by their appearance in one channel but not the others.  The broad
shallow flux enhancement between 1155 and 1165 \AA\ in the LiF1B channel is 
due to the optical anomaly involving the microchannel-plate grid wires as
noted by Sahnow et al. (2000a).

\subsection{Hubble Space Telescope Observations}

The archival Goddard High Resolution Spectrograph (GHRS) observations
of 3C\,273 used in this work include data taken with the G160M
first-order grating covering the wavelength ranges 1214--1425\,\AA\ and
1521--1572\,\AA, and data taken with the Ech-A echelle-mode grating
covering the wavelength range 1250--1257 \AA.  The G160M observations
include data taken as part of HST programs GTO1140, 3951, 4883 ; the Ech-A
data were taken under program GO5719.  
(Ray Weymann was the principal investigator on all of these 
archival $HST$ datasets.)
The G160M observations were obtained before the installation of COSTAR and 
have been described in a number of papers (e.g., Brandt et al. 1993, 1997; 
Penton et al. 2000b).  The post-COSTAR Ech-A
observations, which include 10 separate exposures totaling 19.6
ksec, are described here for the first time.  

The basic calibration and processing of these datasets followed the method
described by Howk, Savage, \& Fabian (1999), with the exception of
refinements to the wavelength scale described below.  All
observations were taken with the large ($2\arcsec\times2\arcsec$)
science aperture of the
GHRS using focal plane motions to mitigate the effects of
fixed-pattern noise.  For data of relatively low signal-to-noise ($S/N$), the
fixed-pattern noise in the GHRS data is a small contribution to the
total error budget.  While one can solve explicitly for the fixed-pattern noise
spectrum and remove it from the data, we applied this correction only in
cases where the fixed-pattern noise obviously affected absorption lines of
interest to avoid adding extra noise into the data as part of the fixed-pattern
reconstruction procedure. The processed G160M data have $S/N$
$\sim8-14$ per resolution element, with the exception of the observations of
\ion{Si}{4} 1402 which have $S/N$ $\sim$ 23.
The GHRS G160M data have a wavelength-dependent resolution of $\sim19$
\kms\ (FWHM) at 1214 \AA\ and $\sim14$ \kms\ (FWHM) at 1550
\AA.  The Ech-A observations have a resolution of $\sim3.5$ \kms\
 (FWHM) and $S/N$ $\sim$ 14.

The absolute wavelength scale used in the standard GHRS pipeline processing
is accurate to roughly $\pm1$ resolution element.  To
place all of the GHRS data on a self-consistent scale, which we also used to 
set the zero point of the $FUSE$ wavelength scale, we used  a ``bootstrapping''
technique.  For a velocity reference, we adopted the \ion{S}{2}
$\lambda\lambda1250.584,1253.811$ lines observed with the Ech-A grating. These
\ion{S}{2} lines have a wavelength solution accurate to $\pm$3.5 \kms\ and
contain a sharp, narrow component at $v_{LSR} = -6$ \kms\ that coincides with 
the peak of the 
\ion{H}{1} 21\,cm emission along the sight line.  We used the Ech-A \ion{S}{2} 
profiles to bring the G160M observations of the same \ion{S}{2} lines into 
the LSR reference frame.  The G160M
observations of the \ion{S}{2} lines also contain the \ion{Si}{2} 1260.422
\AA\ transition, which we used to register the \ion{Si}{2} lines at
1304.370\,\AA\ and 1526.707\,\AA.  The relative wavelength calibration
was carried out using the ISM and IGM lines in the 
wavelength regions covered by the G160M observations; the only exceptions 
to this approach were the observations centered near the interstellar 
\ion{Si}{4} $\lambda\lambda1393.755, 1402.770$ lines, which we 
tied into this self-consistent scale by assuming the 1402.770 \AA\ line
has a velocity similar to that of the interstellar \ion{C}{4} lines.
(We did not use the stronger \ion{Si}{4} $\lambda1393.755$ transition for
this purpose since it is strongly blended with an IGM Ly$\alpha$ absorber.)

\section{Measurements}

We identified all of the absorption lines in each of the $FUSE$
channels using the atomic line list compiled by 
Morton (1991) and the H$_2$ line lists published by Abgrall et al. 
(1993a,b). We also searched for intergalactic lines of \ion{H}{1}, \ion{O}{6}, 
and \ion{C}{3} corresponding to known Ly$\alpha$ 
absorption lines having rest frame equivalent widths\footnotemark\
W$_r \gtrsim 50$ m\AA\ (i.e., N(\ion{H}{1}) $\gtrsim 10^{13}$ cm$^{-2}$).

\footnotetext{Throughout this work, we use the notation $W_{obs}$ to indicate
the {\it observed} equivalent width of a line and $W_r$ to indicate the 
{\it rest frame} equivalent width of a line.  The relationship between
the two quantities is given by $W_{obs} = (1+z) W_r$, where $z$ is the 
redshift of the absorption. For the Milky Way ISM lines, $W_{\lambda} = 
W_{obs}$.}

\subsection{Interstellar Lines}

We estimated continuum
levels for the interstellar 
atomic and H$_2$ lines 
by fitting spectral regions within a few hundred \kms\ of the 
lines with low-order ($n\le$5) Legendre polynomials.  Continuum-normalized
profiles for a selected set of lines are shown in 
Figure~2 as a function of velocity in the local standard of rest (LSR) 
reference frame\footnotemark.  
\footnotetext{The heliocentric to LSR correction in
the direction of 3C\,273 is  $v_{LSR} = v_{helio}+2.8$ \kms, assuming
a solar motion of 16.5 \kms\ in the direction $l$\,=\,53\degr, 
$b$\,=\,+25\degr (Mihalas \& Binney 1981).}
Only the data from the channel yielding the 
best combination of $S/N$ and resolution are shown. 
Absorption features due to other lines, including H$_2$,
occur within $\pm$300 \kms\ of many of the lines shown.  These nearby lines 
are identified immediately above or below each spectrum.

We measured total line equivalent widths ($W_\lambda$) 
and their associated errors using the 
procedures described by Sembach \& Savage (1992).  The 
equivalent widths of the interstellar atomic lines are tabulated in Table~1.
The two equivalent width entries for each line correspond to the values derived
from the data for the two most sensitive channels (LiF1 and LiF2, or SiC1 
and SiC2).  In general, these independent equivalent widths are in excellent 
agreement.  The 1$\sigma$ errors on the measurements were generally taken 
to be the larger of the formal errors derived from the integration or 10\,m\AA.
This latter value was adopted as a lower limit to account for the possible
fixed-pattern noise contributions that could be present over the $\sim100-150$
\kms\ integration range of the equivalent width measurements (see Table~1).

For the \ion{O}{6} $\lambda1031.926$ line, we split the equivalent 
width integration into two velocity ranges (--160 to +100 \kms\ and +100 to 
+240 \kms) because the \ion{O}{6} profile 
contains a broad, positive velocity wing 
that is not seen in any other atomic species.  The high velocity
absorption cannot be 
easily modeled as a lowering of the QSO continuum at these wavelengths;
on the contrary, its existence appears quite robust to 
reasonable continuum placements derived from adjoining spectral regions.
The total interstellar \ion{O}{6}
$\lambda1031.926$ equivalent width is 450$\pm$15 m\AA. 

Most of the molecular hydrogen lines identified in Figures~1 and 2 are 
readily distinguished from the atomic interstellar and intergalactic lines.  
However, there are numerous cases where weaker H$_2$ lines from rotational
levels $J = 3-4$ may contribute to the observed absorption (see \S4 and \S5.1).
We have therefore listed in Table~2 the observed equivalent widths (or limits) 
for several transitions in the $J=3-4$ rotational levels.  The lines listed 
have $f$-values that span those of the $J=3-4$ lines discussed later in 
conjunction with blending issues associated with the IGM measurements. 
No $J=4$ lines are positively detected in the $FUSE$ data down to a 
$3\sigma$ limit of $\approx30$\,m\AA\ [N(H$_2$) $\sim (2-3)\times10^{14}$ 
cm$^{-2}$]. We list the 3$\sigma$
column density limits derived for the H$_2$ lines in the last column of 
Table~2, assuming a linear curve-of-growth relation between $W_\lambda$ and N.

\subsection{Intergalactic Lines}

Numerous intergalactic absorption lines are present in the $FUSE$ spectra of
3C\,273.  We measured the equivalent widths of these lines in the same manner
as for the interstellar lines (\S3.1).  In Table 3 we list the rest 
wavelengths, observed wavelengths, and observed equivalent widths on both
detectors for each IGM line for which the corresponding 
Ly$\alpha$ equivalent width is greater than 50\,m\AA.  The
entries include values for 
\ion{H}{1} Ly$\beta$ (and Ly$\gamma$ when appropriate), \ion{O}{6}
$\lambda\lambda1031.926, 1037.617$, and \ion{C}{3} $\lambda977.020$ 
for the absorbers at 
$z$ = 0.00338, 0.00530, 0.02497, 0.04898, 0.06655, 0.09012, 0.12007, and
0.14660.  For completeness, we also list the equivalent widths of the 
Ly$\alpha$ lines measured with pre-COSTAR GHRS data in 
Table~3.  These Ly$\alpha$ line strengths and redshifts 
are in good agreement with those 
made by previous investigators (e.g., Penton et al. 2000b).  Comments appended
to each entry provide additional information about possible  
blending with Galactic lines.  We have listed equivalent widths for as many 
\ion{H}{1} lines as are observable down to a limiting equivalent width 
$W_{obs} \sim20-30$ m\AA\ ($3\sigma$).

We plot the continuum normalized IGM absorption lines in Figure~3 as a function
of systemic velocity centered on the redshifts listed in Table~3.  In all but 
two cases, the $z = 0.12007$ absorber and possibly the $z = 0.00338$ absorber, 
there is no significant
metal-line absorption detected.  Confusion with Galactic lines is 
problematic for some of the IGM lines.  We discuss possible blends with 
Galactic features for each system in the following section.

\section{Intergalactic Medium Absorption}

A summary of the IGM column densities for all of the 
absorbers studied is provided in Table~4.  Column densities for \ion{H}{1} 
were derived by fitting a single-component Doppler-broadened curve of growth
to the \ion{H}{1} lines detected in each absorber.  Data points from 
independent channels
were included in the error-weighted fits whenever possible.  We followed the 
procedures outlined by Savage, Edgar, \& Diplas (1990) for calculating 
errors on N(\ion{H}{1}) and $b$(\ion{H}{1}). Limits on N(\ion{O}{6}) and 
N(\ion{C}{3}) are conservative 3$\sigma$ estimates derived from the data 
in the channel having the highest sensitivity.  If data from both the 
LiF1 and LiF2 channels are considered, these limits translate into 
$\sim4\sigma$ estimates.  We now discuss each of the absorbers individually.

\subsection{Virgo Cluster Absorbers ($z=0.00338$ and $z=0.00530$)}
The 3C\,273 sight line passes through the Virgo Cluster.  The two strongest
Ly$\alpha$ absorbers along the sight line occur at velocities 
representative of those of the cluster ($\langle V \rangle \approx 1141\pm60$ 
\kms; $\sigma_r \approx 666$ \kms; Tammann 1972).  We 
detect redshifted Ly$\beta$ absorption associated with 
both clouds at $z = 0.00338$ ($cz \approx 1015$ \kms) and $z = 0.00530$ 
($cz \approx 1590$
\kms).  Previous observations of the Ly$\beta$ lines with $ORFEUS$-$II$
(Hurwitz et al. 1998) confirmed the intergalactic 
identification of the longer wavelength
Ly$\alpha$ lines observed with the $HST$ by Morris and co-workers but 
were of insufficient quality to determine the \ion{H}{1}
column densities of the absorbers to a high degree of confidence.  

The Ly$\alpha$ lines for the two Virgo absorbers have similar equivalent widths
($\sim370$ m\,\AA), but the Ly$\beta$ lines differ in equivalent width
by $\sim90-100$ m\,\AA\ and have very different shapes (Figure 3).  Note that 
the measured depth of the Ly$\beta$ line of the 1590 \kms\ absorber is greater 
than the depth of the Ly$\alpha$ line; this is due to the large
instrumental smearing of the Ly$\alpha$ line resulting from the spherical 
aberration of the pre-corrected $HST$ optics.  There is little 
or no detectable Ly$\gamma$ absorption in the lower velocity cloud
(Table~3), but the 
1590 \kms\ cloud can be seen in absorption through the Lyman series up
to Ly$\theta$ without much confusion by Galactic absorption.
These higher-order \ion{H}{1} lines are identified in Figure~1 and are 
plotted as a function of systemic velocity in Figure~4.

The Ly$\beta$ lines of both Virgo absorbers occur at wavelengths where weak 
Galactic H$_2$ absorption may blend with the lines.  We show the 
Lyman series H$_2$ (6--0) R(3) and P(3) lines in Figure~5 together with 
an assortment of R(3) and P(3) lines in other vibrational bands
having line strengths, $f\lambda$, spanning those of the (6--0) lines.  The top
spectra in the two panels of Figure~5 contain the Virgo Ly$\beta$ lines
with fits to the R(3) and P(3) lines shown as heavy smooth curves.
Nearby R(2) and P(2) lines confirm the velocities of the $J=3$ 
lines relative to the IGM lines.  For the 1015 \kms\ absorber, the 
overlapping H$_2$ R(3) line has an (unblended) 
equivalent width of $\approx42$\,m\AA\
and contributes $\approx25$ m\AA\ to the observed (blended) absorption.
For the 1590 \kms\ absorber, the overlapping 
H$_2$ P(3) line has an equivalent width of $\approx34$\,m\AA\ and 
contributes $\approx 16$ m\AA\ to the observed absorption.

We show the \ion{H}{1} curves of growth for the two Virgo absorbers in 
Figure~6.
The COG results for the 1015 \kms\ absorber differ only 
slightly from those derived 
by Weymann et al. (1995) from profile fitting of the Ly$\alpha$ line 
alone\footnotemark: 
log\,N(\ion{H}{1})$_{COG}$ = $14.40\pm0.10$ versus 
log\,N(\ion{H}{1})$_{Ly\alpha}$ = $14.19\pm0.04$,
and $b_{COG} = 30.3\pm^{4.3}_{3.8}$ \kms\ versus 
$b_{Ly\alpha} = 40.7\pm3.0$ \kms.  
\footnotetext{Unless otherwise indicated, errors quoted in this paper are 
$1\sigma$ (68\% confidence) estimates.}
However, the COG results for the 1590 \kms\ absorber differ greatly from those 
derived previously from Ly$\alpha$ profile fitting:
log\,N(\ion{H}{1})$_{COG}$ = $15.85\pm^{0.10}_{0.08}$ versus 
log\,N(\ion{H}{1})$_{Ly\alpha}$ = $14.22\pm0.07$,
and $b_{COG} = 16.1\pm1.1$ \kms\ versus $b_{Ly\alpha} = 34.2\pm3.3$ \kms. 
Curve-of-growth column densities and $b$-values for the 1590 \kms\ absorber 
were derived both including and excluding the Ly$\alpha$ line.
The slight excess of Ly$\alpha$ equivalent width in the 1590 \kms\
absorber over that predicted by the best-fit curve of growth to the 
\ion{H}{1} lines in the $FUSE$ bandpass (Figure~6) is most likely caused by
an additional low column density absorption observable in Ly$\alpha$ that is
essentially undetectable in the higher order \ion{H}{1} lines.  
The excess is $\lesssim40$\,m\AA.  Including Ly$\alpha$ in the 
curve-of-growth analysis increases the derived $b$-value by $\approx1.5$ \kms\
and decreases the derived column density by $\approx0.08$ dex (see Table~4), 
which is consistent with the column derived solely from the higher order
\ion{H}{1} lines observed by $FUSE$ to within the uncertainties of each 
estimation. Higher resolution Ly$\alpha$ observations are needed to determine 
whether this absorber has a multi-component structure as implied by the 
single-component COG analysis.  The 
extant GHRS data are insufficient to determine the detailed velocity 
structure of the Ly$\alpha$  absorption since the instrumental 
line spread function is a complex convolution of both a narrow core and broad 
wings.

In summary, we find the following relationships between our COG values and 
previous determinations of N and $b$ in the Virgo absorbers:
\begin{equation}
cz \sim 1015~{\rm km\,s^{-1}:}~~~~~
\left({\frac{\log {\rm N}_{COG}}{\log {\rm N}_{Ly\alpha}}}\right)\approx1.6, 
~~~~\left(\frac{b_{COG}}{b_{Ly\alpha}}\right)\approx 0.7
\end{equation}
\smallskip
\begin{equation}
cz \sim 1590~{\rm km\,s^{-1}:}~~~~~
\left({\frac{\log {\rm N}_{COG}}{\log {\rm N}_{Ly\alpha}}}\right)\approx43, 
~~~~\left(\frac{b_{COG}}{b_{Ly\alpha}}\right)\approx 0.5 
\end{equation}

\noindent
Synthetic line profiles constructed using the best fit curve-of-growth results
are overplotted on the 1590 \kms\ absorber \ion{H}{1} lines in Figure~4. The 
synthetic profiles have been convolved with a single-component 
Gaussian instrumental line spread function having a fixed width of
FWHM = 0.075\,\AA. The general agreement with the observed profiles 
is excellent; the minor differences observed could be reduced through the 
application of a wavelength-dependent line spread function.  

Unfortunately, it is not possible to unambiguously associate the observed
absorbers with any particular galaxy within the Virgo Cluster since the 
internal cluster velocity dispersion is sufficiently high that numerous
galaxies could be responsible for the absorption. 
McLin et al. (2001) have found a galaxy with a redshift of 
$1512\pm40$ \kms\ that seems to be a plausible source for the 
1590 \kms\ absorber.

The two Virgo absorbers contain metals.  We detect very weak 
\ion{O}{6} $\lambda1031.926$ absorption in the 1015 \kms\ absorber (Figure~3
and Table~3).  The absorption is present in both LiF channels but is weak. 
The conservative errors of 10\,m\AA\ in both channels account for statistical
noise, continuum placement, and fixed-pattern noise uncertainties. 
The Lyman series H$_2$ (6--0) R(3) absorption
due to the Milky Way ISM is expected to be weak and offset from this 
absorption.  Based on the FUSE data, we estimate 
N(\ion{O}{6}) $\approx (2.1\pm0.8)\times10^{13}$ cm$^{-2}$.  
This is the lowest redshift IGM \ion{O}{6} absorber outside the Local Group.  

There is no obvious \ion{O}{6} associated with the 1590 \kms\ absorber
in the 1037.617 \AA\ line.  We place a $3\sigma$ limit of 
N(\ion{O}{6}) $< 4.8\times10^{13}$ cm$^{-2}$ on the presence of \ion{O}{6}.  
The \ion{O}{6} $\lambda1031.926$ and 
\ion{C}{3} $\lambda977.020$ lines are blended with Milky
Way ISM  absorption (Table~3).  Using the H$_2$ parameters discussed in 
\S5.2.3 to deblend the line, we conservatively estimate that 
N(\ion{C}{3}) $\lesssim 1.2\times10^{13}$ cm$^{-2}$ ($3\sigma$), which 
translates into N(H$^+$) $\lesssim 3.2\times10^{16} Z^{-1}$ cm$^{-2}$ under 
the assumption that all of the carbon is in the form of \ion{C}{3}.  Here,
$Z$ is the metallicity in units of the solar metallicity.
Thus, N(H$^+$)/N(\ion{H}{1}) $\lesssim 4.5 Z^{-1}$.
Williger et al. (2001) detect 
\ion{Si}{3} $\lambda1206.500$ absorption, which confirms that ionized gas
is present.  Additional searches for 
other strong ionized gas lines in the $HST$ bandpass (e.g., 
\ion{Si}{4}, \ion{C}{4}) would help to constrain the ionized gas content of 
both Virgo absorbers.

\subsection{The $z=0.02947$ absorber}
This absorber exhibits very weak Ly$\beta$ absorption present at the 
2-3$\sigma$ level associated with the 139 m\AA\ Ly$\alpha$
absorber.  The ratio of Ly$\alpha$ to Ly$\beta$ can be used to determine
a reliable \ion{H}{1} column density, since an optically thin absorber would
have a Ly$\beta$ equivalent width of $W_\lambda$(Ly$\alpha$) / 7.4 
$\approx 19$ m\AA, which is consistent with the observed value of $22\pm9$
m\AA\ averaged over the two LiF channels.
We find N(\ion{H}{1}) $\approx 13.55$ (Table~4).  Since the 
$b$-value is not constrained by the Ly$\alpha$/Ly$\beta$ equivalent width
ratio, we fit a single Voigt profile to the GHRS Ech-A Ly$\alpha$ profile
shown in Figure~3 to estimate $b = 38.7\pm1.7$ \kms.  This $b$-value is 
similar to the value of 36 \kms\ suggested by Morris et al. (1991)
using the  pre-COSTAR GHRS G160M data but is substantially smaller than the 
value of $61\pm8$ \kms\ found by Penton et al. (2000b) with the same G160M
dataset. 

To the best of our knowledge, the Ech-A profile shown in Figure~3 is the 
highest resolution absorption profile
available for any IGM Ly$\alpha$ line at low-moderate redshift ($z \lesssim
1.5$). 
Unless the gas is cold ($T \lesssim 300$ K), which seems unlikely, 
the GHRS Ech-A data fully
resolve the \ion{H}{1} absorption.  No more than $\sim10$\% of the total 
column density can be in any single component with $T < 1000$\,K.
There is noticeable structure within the 
profile; binning the data by a factor of 2 (to an
effective resolution of about 7 \kms\ FWHM) suggests that there may be both 
a narrow ($b \sim 25$ \kms) core containing about 30\% of the column
and a broader ($b \sim 50$ \kms) component containing about 70\% of the 
column. This has important implications, since the broad component could have
a temperature as high as $\sim1.5\times10^5$\,K if the broadening is thermal 
in nature.  (Even if only a single component is fit, the temperature could
be as high as $\sim10^5$\,K.)
Alternatively, if the broadening is non-thermal, 
there must be significant bulk or turbulent motions within the absorber.
Higher $S/N$ observations of the Ly$\alpha$ line at comparable
resolution with STIS could be used to check the validity of this claim and to
search for metal-line diagnostics of $10^5$\,K gas (e.g., \ion{C}{4}) if the 
gas is predominantly collisionally ionized.
No \ion{C}{3} or \ion{O}{6} absorption is detected in the $FUSE$ spectra
of this absorber.  

\subsection{The $z=0.04898$ absorber}
This absorber has an observed
Ly$\alpha$ equivalent width of $\approx126$ m\AA, and 
like the $z=0.02947$ system, the absorption is confirmed to be optically thin
by the absence of a strong Ly$\beta$ absorption.
There is a hint of Ly$\beta$ absorption in the LiF1 data (Figures 1 and 3), 
but the observed wavelength of 1075.95\,\AA\ is not covered by the LiF2 
channel. The \ion{O}{6} lines
fall in the $FUSE$ LiF wavelength coverage gap, so the only measurements 
available come from the lower $S/N$ SiC data (Table~3).  The \ion{C}{3}
line in this absorber falls in the wing of Galactic Ly$\beta$ absorption
and is not observable.
  
\subsection{The $z=0.06655$ absorber}
This is the fourth strongest Ly$\alpha$ absorber along the sight line,
with an observed equivalent width of $\approx 312$ m\AA.  The 
corresponding Ly$\beta$ 
absorption is recorded only in the LiF2 channel because it falls in the 
SiC1/LiF1 wavelength gap.  The line is partially
blended with the Galactic H$_2$ (1--0) P(1) line (Figure~3).  Ly$\gamma$ 
occurs near 1037.3\,\AA\ and is blended with Galactic \ion{C}{2} and H$_2$ 
absorption.  There is no detectable \ion{C}{3} or \ion{O}{6} in this 
absorber.

\subsection{The $z=0.09012$ absorber}
This absorber has a very shallow, irregularly shaped Ly$\alpha$ line extending
over nearly 200 \kms.  There may be very weak Ly$\beta$ absorption at the
2-3$\sigma$ level consistent with an optically thin cloud.  There is no 
detectable \ion{O}{6} absorption.  \ion{C}{3} is blended with the Galactic 
H$_2$ (3--0) R(2) line, which we modeled; we find no residual \ion{C}{3}
down to a level of $<30$ m\AA\ (3$\sigma$). 

\subsection{The $z=0.12007$ absorber}

This absorber has a modest \ion{H}{1} column density, $\log N$(\ion{H}{1}) = 
13.61, and significant detections of \ion{O}{6} and \ion{C}{3}.  
The metal-line absorption is weak [$W_{obs}{\rm (O~VI~\lambda1031.926)} 
\approx 30$ m\AA], but is present in both $FUSE$ channels (Table~3) as well
as in the weaker $\lambda1037.617$ line.
The \ion{C}{3} line falls near a blend of Galactic H$_2$ and IGM Ly$\beta$ 
absorption at a redshift $z=0.06655$ (see Figure~3). 

We find N(\ion{O}{6}) $\approx$ $(2-3)\times10^{13}$ cm$^{-2}$, which is 
smaller than the columns for other \ion{O}{6} absorbers detected with $FUSE$
in the spectrum of PG~0953+415
(see Savage et al. 2001a).  We find N(\ion{H}{1})/N(\ion{O}{6})
= $1.7\pm^{0.6}_{0.8}$, 
indicating that the gas is essentially fully ionized if 
the \ion{H}{1} and \ion{O}{6} are co-spatial.  For an \ion{O}{6} ionization
$O^{+5}$/O $\le 0.2$, the observed value of 
N(\ion{O}{6}) = 
$2.4\times10^{13}$ cm$^{-2}$ implies N(H$^+$) $\ge$ $1.6\times10^{17}\,Z^{-1}$ 
cm$^{-2}$, where $Z$ is the metallicity of the gas\footnotemark.  Thus, 
N(H$^+$)/N(\ion{H}{1}) $\gtrsim 4000\,Z^{-1}$. 
Additional ultraviolet data for other ionized species (e.g., \ion{C}{4}, 
\ion{N}{5}) would be valuable in determining the ionization and 
metallicity of the gas. 
\footnotetext{The peak \ion{O}{6} ionization fraction of $\sim 0.2$ 
occurs in collisional ionization equilibrium at a temperature of 
$\sim3\times10^5$~K (Sutherland \& Dopita 1993).  This ionization fraction
is also rarely exceeded if the gas is photoionized under conditions that might
be found in the low-redshift IGM.}

The nearest known galaxy to the sight line with a redshift within a few
hundred \kms\ of the absorber lies at a projected distance of 
$\sim2.5$ 
Mpc from the 3C\,273 sight line assuming H$_0$ = 65 \kms\ Mpc$^{-1}$
and has a redshift $z=0.12067$
(Morris et al. 1993).  McLin et al. (2001) have also found a galaxy at 
$z = 0.12026$ that is $\sim$16\arcmin, or $\sim 2.6$ Mpc, from the sight line.

\subsection{The $z=0.14660$ absorber}
The Ly$\alpha$ line for this absorber occurs at a wavelength of 1393.89\,\AA\
and is blended with Galactic \ion{Si}{4} $\lambda1393.755$ absorption.  Our 
estimate of the observed
equivalent width of the line ($355\pm20$\,m\AA) 
is in good agreement with the value of 331\,m\AA\ derived by Morris et 
al. (1991) but differs substantially from the value of 
216\,m\AA\ quoted by Savage et 
al. (1993)\footnotemark.
\footnotetext{The lower equivalent width quoted by Savage et al. 
(1993) resulted
from a minor mathematical error in the scaling of the deblended IGM and 
Milky Way absorption features.  Figure~7 in that paper can be compared to our 
revised deblending shown in Figure~7 of this paper.  In both the 
present paper and in the Savage et al. (1993) article, 
this error does not affect the 
conclusions regarding the column density or velocity of \ion{Si}{4}, since that
information is based in both studies on the 1402.770\,\AA\ line.}
We re-derived the Ly$\alpha$ profile  by scaling the observed optical 
depth of the ``clean'' Galactic \ion{Si}{4} $\lambda1402.770$ 
line by a factor of 2, subtracting this
scaled optical depth from the observed optical depth 
of the \ion{Si}{4}\,+\,Ly$\alpha$ blend, and 
recalculating the absorption profile of the residual (e.g., IGM Ly$\alpha$)
absorption.  We show our decomposition of the absorption profile in 
Figure~7. 

We detect Ly$\beta$ absorption in this absorber with an equivalent width 
consistent with an optically thin cloud.  
This is the third strongest \ion{H}{1} absorber along the 3C\,273 sight line.  
There are
no detectable \ion{O}{6} or \ion{C}{3} absorptions associated with the 
intergalactic \ion{H}{1} at this redshift.

\section{Milky Way Interstellar Absorption}
The 3C\,273 sight line passes through the ISM of the Galactic
disk and halo.  In this high latitude direction ($l = 289.95\degr, 
b = +64.36\degr$), Galactic rotation does not 
strongly affect the radial velocity of the interstellar gas; for $d < 10$ kpc,
the expected LSR velocities of a smoothly distributed medium corotating with 
the Galactic disk lie between 0 and --8 \kms.
The most notable features along or near the sight line 
include radio continuum Loops~I and IV (Berkhuijsen 1971) and the North 
Polar Spur, which is seen in both 21\,cm emission (Colomb, Poppel, \& Heiles 
1980; Heiles et al. 1980) and in X-ray emission (McCammon et al. 1983;
Snowden et al. 1995).  Descriptions of these features and an overview of the 
ultraviolet and optical absorption along the sight line are given by Savage 
et al. (1993).  Here, we summarize some of the new information about the 
interstellar gas obtained from the $FUSE$ spectra of 3C\,273.

\subsection{Kinematics}

The neutral and low ionization interstellar gas toward 3C\,273 occurs in
two main groupings of clouds centered at $v_{LSR} \approx -15 $ and +23 \kms,
as derived from high resolution GHRS observations of the \ion{S}{2} 
$\lambda\lambda$1250.584, 1253.811 lines and $FUSE$ observations of
\ion{P}{2} $\lambda1152.818$, \ion{Si}{2} $\lambda1020.699$, and \ion{Fe}{2}
$\lambda1121.975, 1125.448, 1143.226, 1144.938$.  
These absorption features are shown in Figure~8 along with the \ion{H}{1}
21\,cm emission profile obtained with the NRAO 140-foot telescope.  
The ultraviolet absorption in the neutral and singly ionized species occurs 
at velocities for which \ion{H}{1} 21\,cm emission is also detected.
The $-15$ \kms\ grouping is the dominant contributor to both the 
21\,cm emission and the ultraviolet absorption.
The maximum \ion{H}{1} emission occurs at $v_{LSR} \approx -6$ \kms,
the velocity of the 
strong, sharp absorption feature visible in the \ion{S}{2} $\lambda$1250.584
line shown near the bottom of Figure~8.

The \ion{S}{2} profiles observed at high resolution with the GHRS indicate that
the velocity component distributions of the low ionization 
groups of clouds are complex.
The negative velocity grouping contains at least three components, and the 
positive velocity grouping contains at least two components.  Typical 
$b$-values of Gaussian components fit to the profiles are $\sim3-8$ \kms.
  The profiles are sufficiently complicated that it is difficult to obtain a 
  unique solution to the various component parameters; this is particularly 
  true for the negative velocity (--15 \kms) group, which has a trailing 
  absorption shoulder that is difficult to describe as a sum of discrete 
  Gaussian components (see Figure~8).
Comparisons of the apparent column density profiles (next section)
for the \ion{S}{2} lines shows, however, that there is unlikely to be any 
unresolved saturated structure within the lines, except possibly in the 
narrow core at --6 \kms.

The overall shape  of the absorption lines is consistent from species to 
species, with the possible exception of the \ion{Ar}{1} 
$\lambda\lambda1048.220, 	1066.660$ 
lines.  The \ion{Ar}{1} profiles consist of a two-component structure, 
but the cloud group centroids are separated by only $\sim33$ \kms, 
compared to $\sim38$ \kms\ for the ``dominant'' ion lines of \ion{Si}{2},
\ion{P}{2}, \ion{S}{2}, and \ion{Fe}{2}.  This appears to be due primarily to
a shift of the redward \ion{Ar}{1} cloud grouping to slightly less positive 
velocities compared to the other species.  Such an offset could be due to 
differences in the ionization of the components within the positive velocity
clouds.  For example, if the higher velocity components in the group contain
a significant amount of ionized gas, then \ion{Ar}{1}, which is found 
predominantly in the neutral gas and is particularly susceptible to 
photoionization compared to other neutral species (Sofia \& Jenkins 1998), 
may have a different profile shape than the singly ionized species.

  The H$_2$ absorption along the sight line occurs near the velocities of the 
  positive velocity cloud group seen in the neutral and singly ionized atomic 
  lines.  The H$_2$ is centered on an LSR velocity of +16 \kms, slightly less
  than the velocity of the low ionization lines (see Figure~8).  There is 
  little detectable H$_2$ absorption near $-15$ \kms.  Thus, the primary 
  molecular absorption along the sight line does not coincide with the peak 
  atomic column density concentration (in velocity) along the sight line.
This runs somewhat contrary to 
expectations and indicates that it is not always possible to use the 
peak of the \ion{H}{1} 21\,cm emission to set the zero-point of the velocity 
scale for the H$_2$ absorption observed with $FUSE$.

The profiles of the doubly ionized species observed along the sight line 
(\ion{S}{3}, \ion{Fe}{3}) span velocities similar to those 
covered by the singly ionized lines (see Figure~2).  
The \ion{S}{3} and \ion{Fe}{3} lines are smoother than their singly ionized
counterparts, indicating that these ions may trace both the 
structured absorption as well as a more turbulent medium. Because of its great
strength, the \ion{C}{3} $\lambda977.020$ 
line is broader than the \ion{S}{3} or \ion{Fe}{3} lines.  It has an
extent (FWHM $\approx 180 $\kms) comparable to or slightly larger than
that of the strong \ion{C}{2} $\lambda1036.337$ line.  The great 
extent of these lines implies the existence of a low column density
medium that is highly turbulent.

The \ion{O}{6} $\lambda\lambda1031.926, 1037.617$ lines are broad and span 
a velocity range from $-100$ to $+100$ \kms.  The 1031.926\,\AA\
line also reveals a broad, shallow wing of absorption between $+100$ and 
$+240$ \kms\ that is not detected in other high or low ionization
species, including \ion{C}{2}, \ion{C}{3}, \ion{C}{4}, \ion{Si}{4}, 
and \ion{N}{5} (Sembach et al. 1997).  The feature cannot be redshifted 
\ion{H}{1} Ly$\beta$ at $z \approx 0.00638-0.00685$ because there is no
corresponding Ly$\alpha$ absorption at this redshift.
Broad shallow interstellar \ion{O}{6} absorption features are seen
in several other directions through the Milky Way halo (e.g., toward Mrk~1383,
PKS~2005-489;  Savage et al. 2001b).   Toward 
3C\,273, the wing contains roughly 13\% of the total \ion{O}{6} equivalent 
width. The positive 
velocity wing of the \ion{O}{6} $\lambda1037.617$ line is blended with 
absorption by the nearby Lyman H$_2$ (5--0) R(1) line.  The \ion{O}{6} 
lines do not contain any unresolved saturated component structure.

\subsection{Column Densities and Elemental Abundances}

We computed column densities for the interstellar species observed 
toward 3C\,273 by fitting single-component Doppler-broadened curves of 
growth to the equivalent widths listed
in Table~1 and by calculating the apparent optical depths (AOD) of 
the lines (Savage \& Sembach 1991). 
We list the column density results in Table~5 for several species.  
Values  are listed for the two groups of components near --15 and +23 \kms\
 as well as for the entire sight line.  The velocity ranges
considered were similar to those listed in Table~1 for the individual lines.  

The apparent column density obtained by integrating the column density 
profiles over velocity is given by 
\begin{equation}
{\rm N_a} = \int {\rm N_a}(v) dv
= \frac{3.768\times10^{14}}{f\lambda} \int \tau_a(v) dv~~~~{\rm (cm^{-2})}
\end{equation}

\noindent
where $\tau_a(v)$ is the apparent optical depth of the line (equal to the 
natural logarithm of the estimated continuum divided by the observed intensity)
at velocity $v$ (in \kms), $f$ is the oscillator strength of the 
line, and $\lambda$ is the wavelength of the line (in \AA).  

\subsubsection{Neutral and Low Ionization Gas}

The total amount of neutral (\ion{H}{1}) gas along the 3C\,273 sight line
can be estimated in several ways.  Two independent estimates are provided in 
Table~5.  The first of these is an estimate based upon an integration of the 
21\,cm emission profile shown in Figure~8 over the velocity ranges from $-50$ 
\kms\ to $+10$ \kms\ and from $+10$ \kms\ to $+60$ \kms\ under the assumption 
that the emission is optically thin (see eq. 3-38 in Spitzer 1978).  The total
resulting column density is log\,N(\ion{H}{1}) = 20.23.  The second estimate
is based upon a profile fit to the radiation damping wings of the Ly$\beta$
line shown in Figure~1.  The resulting column density is log\,N(\ion{H}{1})
= $20.20\pm0.05$.  Both results are consistent with each other and with the 
strengths of the higher order Lyman series lines observed in the $FUSE$ 
bandpass.  The higher order lines lie on the flat part of the curve of growth
and are often contaminated by metal lines, molecular hydrogen lines,
and geocoronal airglow emission.

Comparisons of the apparent column density profiles for each species revealed 
that many of the low ionization lines contain unresolved saturated structures 
at the resolution of the $FUSE$ data.
The weak-line apparent column density integrals 
for the \ion{Fe}{2} lines indicated that lines having apparent optical depths 
$\tau_a \lesssim 1$ yield the same integrated column density as 
derived from the COG.  Thus, the AOD results for species having weak lines 
of comparable optical depth
(like \ion{P}{2} $\lambda1152$) should be reliable provided that the velocity
structure is similar to that of \ion{Fe}{2}.
To help constrain the $b$-values derived from the COG
for \ion{Si}{2} and \ion{Fe}{2}, we used the equivalent width measurements
reported by Savage et al. (1993) for several strong lines of these species 
observed with the $HST/GHRS$ (see Table~5 notes).  
The \ion{S}{2} results derived 
here from high-resolution GHRS Ech-A data are much more accurate than previous 
values derived from the G160M data; we find a sight line total column density
of $\log N$(\ion{S}{2}) = $15.46\pm0.06$, compared to 
the value of $15.60\pm^{0.24}_{0.15}$ derived by Savage et al. (1993).

A standard uncertainty in deriving accurate estimates of the relative 
elemental abundances is determining to what extent ionization 
affects the column densities of the different ionization stages of each 
element observed.  
The relative gas-phase abundance of elements ``$X$'' and ``$Y$'' is usually 
written as
\begin{equation}
{\rm [X/Y]} = \log {\rm N}(X^i) - \log {\rm N}(Y^j) - ({\rm A}(X)-{\rm A}(Y)),
\end{equation}
where $X^i$ and $Y^i$ are usually the neutral or singly ionized forms of $X$
and $Y$, and A($X$) and A($Y$) are the cosmic (solar) abundances of $X$ and 
$Y$ on a logarithmic scale where A(H) = 12.00. In most abundance studies, 
the total \ion{H}{2} column density is unknown, so it is often assumed that 
singly ionized species trace mostly neutral (\ion{H}{1}) gas and little 
ionized ({\ion{H}{2}) gas.  

Using $FUSE$ observations of both singly and doubly ionized atoms of the same 
element, we can check the validity of the above assumption and its
impact on the elemental abundance estimates.  The doubly ionized forms of 
Fe and S are not present in appreciable quantities
in \ion{H}{1} regions (IP$_{\rm Fe\,II} = 16.18$ eV and 
IP$_{\rm S\,II} = 23.33$ eV)\footnotemark.  
\footnotetext{Ionization potentials in this paper are from Moore (1970).}
We find 
N(\ion{Fe}{3})/(N(\ion{Fe}{2})+N(\ion{Fe}{3})) $\approx 0.18$ and 
N(\ion{S}{3})/(N(\ion{S}{2})+N(\ion{S}{3})) $\approx 0.13$. Thus, the numbers 
of doubly ionized atoms of Fe and S are small compared to their singly
ionized counterparts.   

For a warm photoionized medium characteristic of the ``Reynolds Layer''
having $n_e \sim 0.1$ cm$^{-2}$ and $T\sim10^4$\,K (Reynolds 1993),
the predicted ionization fractions of \ion{Fe}{2} and \ion{Fe}{3} are 
$\sim0.5$ (Sembach et al. 2000a).  Thus, the relative column densities
of \ion{Fe}{2} and \ion{Fe}{3} imply that approximately 36\% of the Fe
along the sight line is in the warm ionized medium (WIM) 
if there are no other sources of 
ionization along the sight line.  This percentage is a factor of about 
1.5 higher than that derived for the sight line to vZ\,1128 ($l = 42.5\degr, 
b = 78.7\degr$; Howk, Sembach, \& Savage 2001a). However, the predicted 
ionization
fractions of \ion{S}{2} and \ion{S}{3} are 0.80 and 0.20, respectively,
implying that most of the S ($\sim65$\%) is in the WIM if no other
ionization sources are present.  This leaves only $\sim35$\% of the S 
available for the neutral gas along the sight line, a result clearly at
odds with the observed amounts of \ion{H}{1} and \ion{P}{2} and the
implied depletion of S out of the gas-phase that would be required in 
the \ion{H}{1} gas.  This result
strongly suggests that there must be other sources of ionization, such
as shocks, that increase the amount of \ion{S}{3} along the sight line
above that expected from a pure WIM layer.

We now investigate the average degree of elemental incorporation into dust 
grains along the sight line for several elements, considering whenever possible
the ionization stages that are dominant in both \ion{H}{1} and \ion{H}{2} 
regions.  First, we compare Fe, a refractory 
element that is readily 
incorporated into dust grains, to S, an element found primarily in the 
gas-phase (Savage \& Sembach 1996).
We find N({\ion{Fe}{2}+\ion{Fe}{3})\,/\,N(\ion{S}{2}+\ion{S}{3})
= 0.21~(Fe/S)$_\odot$ , which indicates that Fe is depleted from the gas 
by an amount comparable to that inferred for warm clouds in the Galactic halo 
(Sembach \& Savage 1996).  
Next, we consider (P/S), (Si/S), and (Ar/S); the absence of a measurement
for \ion{Si}{3} implies N(\ion{Si}{2}+\ion{Si}{3}) $\ge$ N(\ion{Si}{2}).
We find N({\ion{Si}{2}+{\ion{Si}{3})\,/\,N(\ion{S}{2}+\ion{S}{3}) 
$\ge 0.20$ (Si/S)$_\odot$,  
N({\ion{P}{2}+{\ion{P}{3})\,/\,N(\ion{S}{2}+\ion{S}{3}) = 
$(0.6-1.7)$\,(P/S)$_\odot$, and N({\ion{Ar}{1})\,/\,N(\ion{S}{2})
$\ge 0.45$ (Ar/S)$_\odot$.  
The nearly solar (P/S) and (Ar/S) ratios strongly suggest that these three 
elements are found predominantly in the gas phase along the 3C\,273 
sight line.   Neither P nor S is usually depleted
by more than a factor of $\sim2$ even in cold clouds (Jenkins 1987), and Ar 
is a noble element.  The lower limit on (Si/S) is about a factor of two
lower than typically found in Milky Way halo clouds, but would probably
increase by roughly this amount if \ion{Si}{3} could be measured.

All of the above abundance ratios are consistent with the abundances found
in warm, diffuse halo clouds (Sembach \& Savage 1996; Savage \& Sembach 1996).
The ratios for the two cloud groupings along the sight line are 
similar, suggesting that in both cases the absorption occurs primarily in
regions where there the grain mantles have been reduced and the grain
cores exposed.  Furthermore, there are few differences between
the ratios derived under the assumption that all of the singly ionized 
atoms occur in \ion{H}{1} regions and the ratios derived from a combination
of singly and doubly ionized species.  This implies that dust in the warm, 
ionized gas along this sight line has  properties similar to the dust in 
the warm neutral gas.  Finally, we note that  
N(\ion{S}{2})/N(\ion{H}{1}) = $0.98\pm0.16$ (S/H)$_\odot$, where 
N(\ion{H}{1}) is derived from the damping wings of the \ion{H}{1} Ly$\beta$ 
profile shown in Figure~1 (see also Table~5). Even after accounting for 
some \ion{S}{2} in ionized regions along the sight line, it is unlikely
that S is depleted from the gas by more than a factor of $\sim2$.
We summarize the ion ratios and elemental abundance results in Table~6.

\subsubsection{High Ionization Gas}

A comparison of the \ion{O}{6} apparent column density profiles, $N_a(v)$, 
derived from the LiF1 data for the 1031.926\,\AA\ and 1037.617\,\AA\ lines
is shown in the top left panel of Figure~9.  We also plot the 1031.926\,\AA\
profile for the LiF2 channel.
The good agreement of the $N_a(v)$ profiles indicates that there are no 
unresolved saturated structures within the lines, which implies that the 
observed 
profiles are valid, instrumentally-smoothed representations of the true 
column density per unit velocity.  Furthermore, the broad positive velocity
wing is present in all of the profiles. The width
of the main \ion{O}{6} absorption (FWHM $\sim100$ \kms) is governed by 
thermal broadening, turbulent motions, and component structure.  For gas at 
$T = 3\times10^5$\,K, the temperature at which \ion{O}{6} peaks in abundance 
in collisional ionization equilibrium (Sutherland \& Dopita 1993), a 
single component has $FWHM \approx 30$ \kms.  An upper limit to the gas 
temperature, $T_{max} < 3.5\times10^6$ K is implied by the width of the 
\ion{O}{6} line core.  However, it is unlikely that the gas is this 
hot since the fractional abundance of \ion{O}{6} at such high temperatures
is very low ($O^{+5}/O < 2\times10^{-4}$).

The maximum observed optical depth of the \ion{O}{6}
$\lambda1031.926$ absorption is $\tau_a^{max} \approx 1.9$, so the true 
optical depth at line center is likely to be modest given the great breadth 
of the line compared to the instrumental resolution.  We can rule out the 
presence of strong narrow components ($\tau_0 > 3$; $T < 5\times10^4$\,K; 
FWHM $< 12$ \kms) in the N$_a(v)$ profiles based on the good 
agreement between the N$_a(v)$ profiles for the weak and strong lines 
(see Figure~9).   This implies
that the \ion{O}{6} is indeed tracing primarily hot ($T \sim 10^5-10^6$\,K) 
gas.

We derive an \ion{O}{6} column density of $\log N$(\ion{O}{6}) = 
$14.73\pm0.04$
($-160$ to $+100$~\kms) using the apparent column density profiles of the 
1031.926\,\AA\ and 1037.617\,\AA\ lines.  
If we include the positive 
velocity wing in the $FUSE$ measurement, we find $\log N$(\ion{O}{6}) = 
$14.77\pm0.04$  ($-160$ to $+240$ \kms).  
The \ion{O}{6} column density derived from $FUSE$ 
can be compared to $\log N$(\ion{O}{6})  = 
$14.85\pm^{0.10}_{0.15}$ derived from $ORFEUS$-$II$ observations having 
$\sim100$ \kms\ resolution and $\log N$(\ion{O}{6}) $>$ 14.3 derived from 
$HUT$ spectra having $\sim1000$ \kms\ resolution 
(Davidsen 1993).
Neither of these previous observations could resolve the weaker \ion{O}{6}
$\lambda1037.617$ line from neighboring absorption lines.  
To date, this 
is the highest column density of \ion{O}{6} measured along any sight line 
through the Milky Way halo.

We show the \ion{O}{6} apparent column density profile together with 
scaled profiles for  \ion{N}{5}, \ion{C}{4}, and \ion{Si}{4} from 
Sembach et al. (1997) in the other three panels of Figure~9.  The integrated
high ion column density ratios between $-100$ and $+100$~\kms\ are:
\begin{equation}
\frac{\mbox{N(\ion{O}{6})}}{\mbox{N(\ion{N}{5})}}   = 7.2\pm1.1;~~~~
\frac{\mbox{N(\ion{O}{6})}}{\mbox{N(\ion{C}{4})}}   = 1.7\pm0.2;~~~~
\frac{\mbox{N(\ion{O}{6})}}{\mbox{N(\ion{Si}{4})}}  = 8.9\pm1.1
\end{equation}

\noindent
The \ion{O}{6}, \ion{N}{5}, and \ion{C}{4}
high ion $N_a(v)$ profiles trace each other 
remarkably well, indicating that the three species arise in similar regions 
along the sight line.  The velocity extents of the primary absorption are 
similar, and the profiles have similar shapes.  The notable exception to 
this is the \ion{Si}{4} excess at $-7 \le v_{LSR} \le +25$ \kms\ compared
to the other profiles.  \ion{Si}{3} has a lower ionization potential than
the other highly ionized species (IP$_{\rm Si\,III}$ = 33.5 eV), so the 
abundance of \ion{Si}{4} is therefore much more sensitive to photoionization 
by starlight than the abundances of \ion{C}{4}, \ion{N}{5}, and \ion{O}{6}.
The velocities of the \ion{Si}{4} excess correspond to those of the 
positive velocity grouping of low ionization species and the H$_2$ along the 
sight line.  

At the higher positive velocities 
of the broad \ion{O}{6} absorption ($+100$ to $+240$ \kms), 
we place the following 
limits ($3\sigma$) on the high ion column densities: 
$\log N$(\ion{N}{5}) $<$ 13.3, $\log N$(\ion{C}{4}) $<$ 13.0, 
and $\log N$(\ion{Si}{4}) $<$ 12.9.  Combined with 
$\log N$(\ion{O}{6}) = $13.71\pm0.05$, these values yield:

\begin{equation}
\frac{\mbox{N(\ion{O}{6})}}{\mbox{N(\ion{N}{5})}}   \ge 2.6;~~~~
\frac{\mbox{N(\ion{O}{6})}}{\mbox{N(\ion{C}{4})}}  \ge 5.1;~~~~
\frac{\mbox{N(\ion{O}{6})}}{\mbox{N(\ion{Si}{4})}} \ge 6.5
\end{equation}

\noindent
It is not possible to place a limit on \ion{C}{3} $\lambda977.020$ at these
velocities because of confusion with a strong IGM 
Ly$\gamma$ line at $z=0.00530$.

\subsubsection{Molecular Gas}
Molecular hydrogen lines are found throughout the $FUSE$ spectrum of 3C\,273.  
Although they are not the primary focus of this investigation, it is 
nonetheless necessary to accurately characterize the column densities in 
several of the rotational levels so that the impact of these lines on the 
absorption observed for other species can be assessed.  In our census of the 
IGM absorbers along the sight line, the $J=3$ rotational level lines are 
the most frequent Galactic absorption features that pose blending problems.  

We modeled the H$_2$ lines in the $J=0-3$ rotational levels by measuring the 
equivalent widths of lines free of contamination by other absorption features 
(see Figure~1).  The lines of each level
were placed on a single-component COG to estimate a  
column density and $b$-value for the molecular gas.  
Using this information, we 
constructed a synthetic H$_2$ spectrum and 
compared it to the data.   We then varied 
the $b$-value and column densities of the individual levels to determine 
the robustness of the fit to the data and to estimate errors on the 
column densities. Examples of the fits obtained for several 
$J=2-3$ lines are shown in Figure~5,
including a few cases where the lines overlap IGM absorption lines.  
Whenever 
we assessed the blending of the H$_2$ and IGM lines, we used H$_2$ lines 
having line strengths, $f\lambda$, bracketing those of the blended lines
(see Table~2)

We estimate a total sight line H$_2$ column density of 
$\log N$(H$_2$) $\approx 15.71$ in rotational levels $J=0-4$.
We find log\,N$_J$(H$_2$) = $15.00\pm0.30$, $15.48\pm0.18$, $14.76\pm0.12$,
$14.73\pm0.12$, and $<14.40$ ($3\sigma$) for $J=0,1,2,3$, and 4, respectively
(Table~5).
The inferred Doppler parameter is $6\pm2$ \kms, and 
the kinetic temperature derived from the relative populations of the 
$J=0,1$ levels is loosely constrained at $T_{01}\approx157\pm100$\,K. 
 Comparing the H$_2$ column to the total 
\ion{H}{1} from 21\,cm emission yields f(H$_2$) = 
2N(H$_2$)/[N(\ion{H}{1})+2N(H$_2$)] $\approx$ $6\times10^{-5}$, typical for 
low-density sight lines through the Milky Way halo (Shull et al. 2000b).
For any reasonable values of N(CO)/N(H$_2$), the CO (A--X) bands 
in the $HST$ bandpass will not be observable.  Thus, the primary 
source of information about molecular gas along the sight line is the H$_2$ 
absorption in the $FUSE$ bandpass.

\section{Discussion}

\subsection{Intergalactic Medium}

A long-standing goal of cosmology is to make an accurate assessment of the 
amount of baryonic material in the Universe.  Using $HST$, it has been 
possible to survey \ion{H}{1} Ly$\alpha$ absorption along numerous 
sight lines.  The 3C\,273 sight line is one of
the best studied directions.  Although most of the Ly$\alpha$ absorbers are 
weak ($\log N$(\ion{H}{1}) $<$ 14), several absorbers have higher
\ion{H}{1} column densities, and the detection of \ion{O}{6}
in two systems ($z=0.00338, 0.12007$) implies very large values of 
N(H$^+$)/N(\ion{H}{1}).  Standard photoionization models of low-$z$ 
clouds also suggest that a significant fraction of the baryonic material 
may be contained in the form of ionized gas at low redshift 
(e.g., Shull, Penton, \& Stocke 1999).  Determining whether the gas is 
photoionized or collisionally ionized remains an outstanding problem (see,
e.g., Savage et al. 2001a).

Our $FUSE$ observations of Ly$\beta$ absorption in 8 of the IGM clouds yield 
refined column densities and Doppler parameters for these absorbers (Table~4).
Previous column density estimates based on Ly$\alpha$ profile fitting generally
produced reliable results for the weak absorbers along the sight line.  
However, the column density of the 1590 \kms\ Virgo absorber measured 
previously was underestimated because the absorption has a considerably smaller
Doppler parameter than inferred from the profile fitting.  For the 8 absorbers,
we find $\langle b \rangle = 25\pm10$ \kms\ with a full spread of $16 
\lesssim b$(\kms) $ \lesssim 46$.  This range of Doppler parameters is similar
to that inferred from the Ly$\alpha$/Ly$\beta$ ratios for stronger Ly$\alpha$
forest absorbers along low-redshift sight lines (Shull et al. 2000a) and at 
redshifts of 2.0--2.5 (Rauch et al. 1993).  For the best 
constrained case of the 1590 \kms\ Virgo absorber, the $b$-value of 16.1 \kms\
implies $T \lesssim 15,000$ K.  This temperature is likely to be even smaller 
if multiple absorption components are present in the profile as expected (see
\S4.1).

For the intergalactic path to 3C\,273, we expect to find $\sim1\pm1$ \ion{O}{6}
absorber with $W_r \gtrsim 30$ m\AA\ after accounting for line blocking and
assuming $dN/dz \approx 15$ (Tripp, Savage, \& Jenkins 
2000; Savage et al. 2001a).  We find at least one, and probably two, 
\ion{O}{6} absorbers along the sight line.  The $z=0.12007$ absorber has been
confirmed by Williger et al. (2001) with STIS data.  

The \ion{O}{6} absorption
may trace the hot ($T > 10^5$\,K) remnants of hierarchical galaxy formation
in the presence of cold dark matter as predicted by N-body simulations 
(e.g., Cen \& Ostriker 1999; Dav\'e et al. 1999).  Although there has yet been
no direct confirmation that any of the \ion{O}{6} absorbers are collisionally
ionized plasmas at high temperatures, there are some excellent cases
that favor collisional ionization over photoionization unless the absorbers
have very low densities ($n < 10^{-5}$ cm$^{-3}$) and large sizes 
($L >100$ kpc) (e.g., Tripp et al. 2001; Savage et al. 2001a). Alternate
hypotheses for the origin of the \ion{O}{6} absorption include dynamical
heating of gases in groups of galaxies or expulsion of interstellar material 
by starbursts (e.g., Heckman et al. 2001).  There is no prominent group of 
galaxies at the redshift of the $z=0.12007$ absorber, and the closest 
known galaxy has a sufficiently large impact parameter ($\sim2.5$ Mpc assuming
H$_0$ = 65 \kms\ Mpc$^{-1}$;
Morris et al. 1993; McLin et al. 2001) 
that expelled material is unlikely to be a source for the 
absorption.  

The Virgo Cluster is a potential source for the weak \ion{O}{6} 
absorption associated with the $z=0.00338$ absorber.  Extreme ultraviolet
observations of Virgo suggest that the cluster contains a gaseous component
with temperatures of $(0.5-1.0)\times10^6$\,K
(Lieu et al. 1996; Bonamente, Lieu, \& Mittaz 2001), but an upper limit on the 
amount of \ion{O}{6} emission observed by $FUSE$ (Dixon et al. 2001a) 
casts doubt upon the widespread existence of a strongly-emitting
intracluster medium with 
$T<10^6$\,K.  Our absorption-line measurements are orders of magnitude more 
sensitive to $10^5-10^6$\,K gas than these emission measurements, but 
the 3C\,273 sight line lies well off the center of the cluster, where 
the amount of extreme ultraviolet 
emission and hot gas observed at X-ray energies 
above $\sim2$ keV is greatest (see Forman \& Jones 1982).  However, there 
are lower levels of X-ray emission in other regions of the cluster 
(Forman et al. 1979), so it is nonetheless possible that the weak \ion{O}{6} 
absorption toward 3C\,273 at $z=0.00338$ is related to hot gas within 
the cluster.  The preferred Doppler width of the \ion{H}{1} profile for the
absorber derived from the curve of growth suggests a temperature of 
$\lesssim55,000$~K; this value rises to $\lesssim100,000$~K ($3\sigma$) if 
the error estimates on $b$(\ion{H}{1}) are included in the limit.  In either
case, it is likely that the absorber has a multi-phase structure if it 
is collisionally ionized.

X-ray 
absorption-line observations of \ion{O}{7} and \ion{O}{8} toward 
3C\,273 may prove useful for determining whether the \ion{O}{6} absorption
traces hot ($10^5-10^6$\,K), collisionally ionized gas or warm ($10^4-10^5$\,K)
photoionized gas.  Unfortunately, the column densities 
of these species would need to be very large (log~N $>$ 16--17)
to be studied spectroscopically with the $Chandra$ observatory.
The most effective method currently available for studying the low-redshift IGM
is ultraviolet absorption-line spectroscopy.
Further absorption-line observations of other low-redshift quasars with $FUSE$
and STIS will help to quantify the redshift distribution of the \ion{O}{6} 
absorbers and their relationship to the IGM Ly$\alpha$ clouds and galaxies.

\subsection{The Hot Interstellar Medium}
The interstellar \ion{O}{6} absorption toward 3C\,273 provides
new information about the physical processes governing the production of hot 
gas along the sight line.  Of all the high ions accessible in the 
ultraviolet spectral
region, \ion{O}{6} is the best 
diagnostic of hot ($T \sim 10^5-10^6$ K), collisionally ionized 
gas since its abundance in the Milky 
Way ISM is unaffected by photoionization (IP$_{\rm O\,V} = 114$~eV).
\ion{O}{6} has been observed along numerous sight lines in the Milky
Way and Magellanic Clouds (Jenkins 1978; Savage et al. 2000, 2001b;
Howk et al. 2001b), but the relationship of the hot gas absorption
to the absorption traced
by other species (e.g., \ion{C}{4}, \ion{N}{5}, and low ions) has not yet 
been studied in detail.  

We list the integrated ratios of \ion{O}{6} to \ion{N}{5}, \ion{C}{4},
and \ion{Si}{4} for five sight lines in Table~7.  
The ratios N(\ion{O}{6})/N(\ion{N}{5}) and N(\ion{O}{6})/N(\ion{Si}{4}) 
vary by factors of $\sim2-3$ in the sample.  N(\ion{O}{6})/N(\ion{C}{4}) 
is similar along four of the five sight lines 
(3C\,273, Mrk\,509, PKS\,2155-304, and H\,1821+643) but is a factor 
of $\approx3$ lower toward ESO\,141-G55.  Unlike the other sight lines,
the ESO\,141-G55 sight line passes through the inner regions of the 
Galaxy, which may account for some of the difference in 
N(\ion{O}{6})/N(\ion{C}{4}).  Spitzer (1996) noted that the 
N(\ion{O}{6})/N(\ion{C}{4}) ratio along 6 halo and 5 disk sight lines 
observed with $Copernicus$ changed from an average value of 
$\sim6.8$ in the disk to $\sim1.1$ in the halo.  The ionized gas
sampled by the extended sight lines listed in Table~7 probably resides
primarily in the low halo rather than in the disk.

The constancy of N(\ion{O}{6})/N(\ion{C}{4}) as a function of velocity along 
the 3C\,273 sight line strongly suggests that the two ions are located in
similar regions and are produced by similar processes.  Models of turbulent 
mixing layers (TMLs) having post-mixed gas temperatures of 
$\sim(2-5)\times10^5$~K and hot gas entrainment velocities of $\sim25-100$
\kms\ produce more \ion{C}{4} than \ion{O}{6}, typically with 
N(\ion{O}{6})/N(\ion{C}{4}) $\sim$ 0.1--0.8 (Slavin,
Shull, \& Begelman 1993).  Over the same temperature range, 
time-averaged values of N(\ion{O}{6})/N(\ion{C}{4})
in non-equilibrium radiatively cooling gas are typically $\sim7-14$
(Edgar \& Chevalier 1986).  Similarly, models that follow the evolution of 
supernova remnants yield time-averaged values of $\sim5-10$
(Slavin \& Cox 1992).  In very late stages of evolution 
($t \gtrsim 1.3\times10^7$ years), the ratio may fall below 2, but generally
it is much higher (Shelton 1998).  The N(\ion{O}{6})/N(\ion{N}{5}) ratio 
is typically $\gtrsim5-10$ in both types of cooling, but can drop to lower
values in the hotter TMLs.  Thus, the interpretation of the observed ratios 
toward 3C\,273 and the other objects listed in Table~7 requires a combination
of physical processes to produce the high ions 
(see, e.g., Shull \& Slavin 1994).

Sembach et al. (1997) proposed a hybrid model for the hot gas in the 
Loop~IV region of the sky to explain the observed high ion ratios
along several sight lines, including 3C\,273; the 3C\,273 sight line passes 
near the edge of the radio continuum emission and X-ray enhancements 
associated with the North Polar Spur.  They suggested that the high ion
absorption occurs within a highly fragmented medium within the Loop~IV 
remnant or in the outer cavity walls of the remnant.  A similar interpretation
was suggested by Heckman et al. (2001) for the outflowing material in the 
starburst galaxy NGC\,1705. Our observations of 
3C\,273 are consistent with this interpretation and lend additional support
for multiple types of gas along the sight line.  In particular, the high 
velocity absorption wing observed only in \ion{O}{6} (\S5.2.2) indicates 
that there is hot gas at substantially different velocities than the bulk of 
the material along the sight line.

High velocity \ion{O}{6} is seen along many sight lines 
through the Milky Way, but in most cases the \ion{O}{6} absorption can be 
associated with lower ionization stages and appears to be more tightly 
confined in velocity than the absorption wing toward 3C\,273 
(e.g., Sembach et al. 2000b; Heckman et al. 2001).  Broad wings on 
\ion{O}{6} profiles are seen along several other sight lines (Savage et al.
2001b), but the detection toward 3C\,273 is the only one for which additional
information for other highly ionized species is presently available.
The high velocity \ion{O}{6} has N(\ion{O}{6})/N(\ion{C}{4}) $>5.1$,
which is consistent with the expectations for radiatively cooling gas 
in fountain flows or young supernova remnants.  Furthermore, the 
ratios of \ion{O}{6} to \ion{N}{5} and \ion{C}{4} are also consistent 
with the values recently determined by Hoopes et al. (2001) for a 
young remnant in the Small Magellanic Cloud [N(\ion{O}{6})/N(\ion{N}{5})
= $17.8\pm3.1$ and N(\ion{O}{6})/N(\ion{C}{4}) = $7.9\pm0.8$], providing
support for the origin of the \ion{O}{6} in a supernova remnant region.

The broad \ion{O}{6} positive velocity
absorption wing toward 3C\,273 is probably the manifestation of hot gas being 
vented out of the Galactic disk into the halo.  The outflow of hot 
($T \sim 10^6$ K) gas is expected in regions where there are multiple 
supernova explosions; the ``chimneys'' serve as conduits for the hot gas
that may eventually return to the Galactic disk as cool high velocity clouds 
(e.g., Shapiro \& Field 1976;
Bregman 1980; Norman \& Ikeuchi 1989).  In the direction of 3C\,273, 
there are clear signatures that multiple energetic events in the disk have 
produced hot gas.  Radio Loop~I is filled with X-ray emission (Snowden et al. 
1995), Loop IV was probably created by the events that reheated Loop~I 
(Iwan 1980), and the North Polar Spur is prominent in X-ray emission
(Snowden et al. 1995).  
Alternatively, the high velocity wing may trace the tidal debris wakes of 
HVCs moving through the Galactic halo or the remnants of 
infalling or tidally disturbed galaxies (e.g., the Sagittarius dwarf
or Magellanic Stream).  Until additional information can be obtained that 
would provide support for the debris hypothesis, we favor the former 
interpretation as the most likely source of the high velocity \ion{O}{6}.

\subsection{Limits on Interstellar O VI Emission}

Dixon et al. (2001b) have recently reported a serendipitous detection of 
\ion{O}{6} emission from the diffuse ISM of the Milky Way 
in the direction of the Virgo Cluster. 
Observing with the LWRS apertures on $FUSE$ for 11 ksec, they found an 
interstellar \ion{O}{6} $\lambda1031.926$ photon surface 
brightness of $2900\pm700$ ph cm$^{-2}$ s$^{-1}$ sr$^{-1}$ and an intrinsic
line width of $<80$ \kms\ (FWHM) toward M\,87 in the direction $l=284.1\degr, 
b = +74.5\degr$.  
Such a result may be fortuitous, since the inferred \ion{O}{6}
column density they derived from a low $S/N$ absorption-line spectrum of the 
nucleus of M\,87 is only $(1.4\pm0.8)\times10^{14}$ cm$^{-2}$, which is a 
factor of 4 less than the value of $(5.4\pm0.5)\times10^{14}$ cm$^{-2}$
	we measure toward 3C\,273 (about 10.3\degr\ away)
and a factor of $\sim2$ 
less than the high-latitude sight line toward NGC\,5548 
($l=32.0\degr, b = +70.5\degr$; Savage et al. 2001b).
Furthermore, Shelton 
et al. (2001) required nearly 200 ksec of integration time with the 
same $FUSE$ 
LWRS apertures
to detect diffuse emission at a level of $2930\pm700$ 
ph cm$^{-2}$ s$^{-1}$ sr$^{-1}$ in the direction $l=315.0\degr, 
b = -43.1\degr$.  In this latter case, it is possible that a nearby 
interstellar cloud obscures much of the emission from material beyond a 
few hundred parsecs.

During our absorption-line
observation of 3C\,273 through the LWRS apertures, we also
observed nearby regions of sky in the MDRS ($4\arcsec\times20\arcsec$)
apertures.  The LWRS and MDRS apertures are separated by an angle of 
approximately 3.5\arcmin\ on the sky and have a size ratio of 11.25. 
The MDRS aperture provides a filled-slit spectral resolution of 
$\approx0.1$\,\AA\ ($\approx 30$ \kms).  In Figure 10 we show a portion of the 
MDRS data for the LiF1 channel, 
processed and extracted in the same manner as the LWRS data.  
The dashed line in the figure shows an emission line 
having a surface brightness of 15,000 
ph\,cm$^{-2}$\,s$^{-1}$\,sr$^{-1}$ and a width the same as that of the 
absorption toward 3C\,273 (FWHM $\approx 120$ \kms\ -- see \S5.1).  
The two thick 
solid curves overplotted on the data are the profiles expected for 
\ion{O}{6} emission having FWHM = 80 \kms\ and photon surface brightnesses
of 3000 and 15,000 ph\,cm$^{-2}$\,s$^{-1}$\,sr$^{-1}$, comparable to and a 
factor of 5 greater, respectively, 
than the emission observed by Dixon et al. (2001b) toward M\,87.
The 3$\sigma$ upper limit of 15,000 
ph\,cm$^{-2}$\,s$^{-1}$\,sr$^{-1}$ set by 
the 3C\,273 data corresponds to an integrated line flux of $5.4\times10^{-16}$ 
erg\,cm$^{-2}$\,s$^{-1}$ through the MDRS aperture.

We cannot rule out \ion{O}{6} emission at the level observed by 
Dixon et al. toward M\,87, but we find that the factor of 4--5 larger 
\ion{O}{6} column density for the 3C\,273 sight line does not likely have a 
corresponding factor of 5 larger surface brightness of \ion{O}{6} emission. 
The limit for the 3C\,273 sight line is less than the level of diffuse 
interstellar \ion{O}{6} emission 
($\sim36,000\pm17,000$ ph\,cm$^{-2}$\,s$^{-1}$\,sr$^{-1}$) 
reported by Dixon, Davidsen, \& Ferguson (1996) 
toward NGC\,4038 in the direction $l=287.0\degr, b = +42.5\degr$, about 
21\degr\ from 3C\,273.  The observed differences in \ion{O}{6} emission 
strength along the 3C\,273, M\,87, and NGC\,4038 sight lines may result from 
differences in the temperatures of the \ion{O}{6}-emitting gases in the 
three directions.  3C\,273 lies near an X-ray bright region of
the North Polar Spur, whereas M\,87 lies in a region of slightly
lower soft (1/4 keV) X-ray brightness, and NGC\,4038 is in a direction of 
much lower soft X-ray emission (Snowden et al. 1995).

A primary problem with interpreting the presence or absence of \ion{O}{6}
emission along a sight line is the unknown geometry of the emitting gas and 
its relationship to the absorbing gas.  Until now, this problem has been 
confounded by the additional uncertainties incurred when measuring emission
and absorption along different sight lines separated by several degrees or 
more on the sky (see, e.g., Dixon et al. 1996).  The $3\sigma$ limit for 
the ratio of \ion{O}{6} 
$\lambda1031.926$ surface brightness (in energy units) to \ion{O}{6} column 
density toward 3C\,273 
is $I_{\lambda1032}$\,/\,N(\ion{O}{6}) 
$<5.4\times10^{-22}$ erg~s$^{-1}$~sr$^{-1}$, 
under the standard assumption that the \ion{O}{6} absorption and emission 
arise in the same gas.\footnotemark\   
\footnotetext{Note that 1031.926\,\AA\ photons have energies of 
$1.925\times10^{-11}$ erg.}
This can be compared to 
a value of $(4.0\pm2.5)\times10^{-22}$ erg~s$^{-1}$~sr$^{-1}$
toward M\,87 using measures of $I_{\lambda1032}$ and N(\ion{O}{6}) from
Dixon et al. (2001b).

Following Shull \& Slavin (1994, see their Eq. 5), we express the electron 
density in terms of the observable quantities and the electron-impact 
excitation rate coefficient, ${\langle \sigma v \rangle}_e$, to write
\begin{equation}
n_e = \frac{3}{2}~\frac{4\pi}{\langle \sigma v\rangle}_e~\frac{I_{\rm 1032}}{\rm N(\mbox{\ion{O}{6}})} = 2.28\times10^{17}~\exp\left[ {\frac{1.392\times10^5\,{\rm K}}{T}}\right]~\frac{T^{1/2}}{\bar{\Omega}(T)}~\frac{I_{\rm 1032}}{\rm N(\mbox{\ion{O}{6}})}
\end{equation}
\noindent
where $\bar{\Omega}(T)$ is the Maxwellian-averaged
 collision strength for de-excitation of 
the \ion{O}{6} doublet.  The factor of 3/2 accounts for the conversion of 
doublet quantities to values appropriate for the 1032\,\AA\ line, and the 
ratio $I_{\lambda1032}$\,/\,N(\ion{O}{6}) has units of 
erg~s$^{-1}$~sr$^{-1}$.  For $T=2\times10^5$\,K and $T=5\times10^5$\,K,
$\bar{\Omega}(T)$ = 5.65 and 6.31, respectively (see Shull \& Slavin 1994).  
Over this temperature range, we find $n_e \lesssim 0.02$ cm$^{-3}$, which 
yields a limit on the thermal pressure 
P/k = $1.92 n_e T  \lesssim 7700$ cm$^{-3}$\,K for $T \approx 2\times10^5$\,K 
and $\lesssim 19,200$ cm$^{-3}$\,K for $T \approx 5\times10^5$\,K,
assuming a fully ionized gas and $n_{He}/n_H = 0.1$.  Note that if the 
absorption and emission do not arise in the same gas, these
limits would increase if foreground material absorbs the emitted \ion{O}{6}
photons.

Time-dependent ionization models should be used to compute the ratio 
$I_{\lambda1032}$\,/\,N(\ion{O}{6}).
Predicted time-averaged values of $I_{\lambda1032}$\,/\,N(\ion{O}{6}) in 
units of 10$^{-22}$ erg\,s$^{-1}$\,sr$^{-1}$ range from \mbox{$\sim2.0-2.2$}
 for turbulent mixing layers with post-mixed gas temperatures of 
$\sim(2-5)\times10^5$\,K 
(Slavin et al. 1993) to $\sim10$ for conductive interfaces (Borkowski,
Balbus, \& Fristrom 1990).  A hot fountain flow with a mass flux of 
\mbox{$\sim4$~M$_\odot$~yr$^{-1}$} to either side of the Galactic plane is 
predicted 
to have a value of $\sim1.3-3.4$, depending upon whether the cooling is 
isochoric or isobaric (Edgar \& Chevalier 1986).  However, alternate cooling 
flow models predict values in excess of $\sim10$ (e.g., Shapiro \& Benjamin 
1992)\footnotemark.
Note that none of these predictions account for the effects of
foreground absorption on the observed flux of the emitted \ion{O}{6} photons.  
A hybrid model of hot gas production involving both turbulent mixing and 
conduction or radiative cooling of hot ($T \sim 10^6$\,K) gas as suggested
by Sembach et al. (1997) for the 3C\,273 sight line and other directions 
in the Loop~IV region is consistent with the available emission-line data.  
\footnotetext{The factor of $\sim10$ difference compared to the Edgar \&
Chevalier (1986) models arises from the different densities assumed for the 
cooling gas ($10^{-2}$ vs. 10$^{-3}$ cm$^{-3}$).}

\subsection{The Warm Ionized Interstellar Medium}

The WIM is a fundamental component of the ISM that has been studied primarily
through optical emission-line observations of H$\alpha$, [\ion{S}{2}],
and [\ion{N}{2}] (see Reynolds 1993).  To date, there has been little 
high quality absorption-line information available to determine the 
fundamental properties
of the WIM other than a study of the \ion{Al}{3} scale height 
($h_{\rm Al\,III} \sim h_{e^{-}} \sim 1$ kpc; Savage et al.
1990) and a study of the WIM dust composition derived from \ion{S}{3}
and \ion{Al}{3} observations for several sight lines (Howk \& Savage 1999).
This latter study found that that the degree of grain destruction in the 
WIM is comparable to that in the warm neutral medium (see Sembach
\& Savage 1996).  Our observations of \ion{S}{3} and Fe\,{\sc ii-iii} toward 
3C\,273 confirm this result.

The great breadth of the interstellar \ion{C}{3} absorption toward
3C\,273 indicates that
the moderately ionized interstellar gas has a velocity extent comparable 
to or slightly greater than that of the neutral gas.  Beyond $\sim-100$ \kms, 
N(\ion{C}{3})/N(\ion{C}{2}) $>1$ and N(\ion{C}{3})/N(\ion{C}{4}) $\gtrsim3$, 
conditions that are not both satisfied for neutral or diffuse photoionized
regions but are met for post-shocked regions (see Shull \& McKee 1979).

The amount of \ion{S}{3} relative to \ion{S}{2} observed toward 3C\,273 
is not compatible with production in a purely photoionized WIM of the type 
described by Sembach et al. (2000a).  The requirement for other sources of 
ionization is not surprising given the large amount of highly ionized gas 
observed along the sight line.  Comparison of the 3C\,273 WIM results to 
similar measurements for halo sight lines that do not pass through
supernova remnants or other known sources of hot gas would help to 
characterize the general
relationship of the hot and warm ionized components of the 
ISM. A study of the optical line emission  in the  immediate vicinity of the 
3C\,273 sight line may also help to quantify the contributions of the 
various high ion production mechanisms to the \ion{S}{3} absorption.

\section{Summary}
We have observed the ultraviolet-bright quasar 3C\,273 
to study the intergalactic and interstellar
absorption along the sight line.  We briefly summarize the primary results of 
this investigation as follows:

\noindent
1) We searched for Ly$\beta$ and metal-line (\ion{C}{3}, \ion{O}{6}) absorption
in the 8 known IGM Ly$\alpha$ absorbers along the sight line having 
$W_r{\rm (Ly}\alpha) > 50$ m\AA.  The detections of Ly$\beta$ confirm the identifications
of the Ly$\alpha$ lines and  provide refined estimates of the 
\ion{H}{1} column densities and Doppler parameters of the clouds.
We find a range of Doppler parameters, $b \approx 16-46$ \kms.

\noindent
2) We detect \ion{O}{6} absorption in the 1015 \kms\
Virgo absorber at the $2-3\sigma$ level.  This is the lowest redshift 
IGM \ion{O}{6} absorber identified outside the Local Group.  The \ion{O}{6} 
absorption may be associated with hot, X-ray emitting gas in the cluster.  
However, the Doppler width of the \ion{H}{1} absorption suggests that cooler 
gas ($T \lesssim 10^5$~K) is also present.

\noindent
3) Multiple Lyman series \ion{H}{1} lines (Ly$\beta$--Ly$\theta$) in addition 
to Ly$\alpha$ are 
detected in the 1590 \kms\ Virgo absorber.  We revise the previous
\ion{H}{1} column density estimate obtained by Ly$\alpha$ profile 
fitting upward by a factor of $\sim43$ to $\log N$(\ion{H}{1}) =
$15.85\pm^{0.10}_{0.08}$.  {\it This absorber is now known to contain ten times
more \ion{H}{1} than all of the other absorbers along the sight line 
combined.}  The Doppler parameter decreased from a value of 
$\sim34$ \kms\ to a value of $\sim16$ \kms.  This line width implies $T \le
15,000$\,K.  Using a limit on the amount of \ion{C}{3} present, we find 
N(H$^+$)/N(\ion{H}{1}) $<4.5 Z^{-1}$.  This absorber may contain multiple
components.  

\noindent
4) We detect \ion{C}{3} and \ion{O}{6} in the absorber at $z=0.12007$.
The \ion{O}{6} absorption is weak 
[$W_\lambda$(\ion{O}{6} 1031) $\approx 30$ m\AA].
This absorber is predominantly ionized and has N(\ion{H}{1})/N(\ion{O}{6})
$\approx 1.7\pm^{0.6}_{0.8}$, corresponding to N(H$^+$)/N(\ion{H}{1})
$\ge 4000\,Z^{-1}$.  Other absorbers along the sight line have limits on 
N(\ion{H}{1})/N(\ion{O}{6}) ranging from $>0.5$ ($z=0.04898$) to
$>148$ ($z = 0.00530$).
Approximately $1\pm1$ \ion{O}{6} absorber with $W_r \gtrsim 50$ m\AA\ 
is expected for the sight line given current estimates of the number 
density of \ion{O}{6} absorbers at low redshift.  

\noindent
5) We analyzed high-resolution (Ech-A) GHRS observations of the Ly$\alpha$ 
absorption at $z = 0.02947$.  The absorber is optically thin and appears
to consist of a combination of broad ($b \sim 50$ \kms) and narrow
($b \sim 25$ \kms) components.  If thermal broadening dominates the 
former, the temperature of the gas may exceed $\sim10^5$ K.

\noindent
6) We measured strong interstellar \ion{O}{6} in the direction of 3C\,273.
The main absorption spans a velocity range from $-100$ to $+100$ \kms.
The \ion{O}{6}, \ion{N}{5}, and \ion{C}{4} lines have very similar
shapes, with roughly constant  ratios across the --100 to +100 \kms\
velocity range.  There is an enhancement in the ratio of \ion{Si}{4} to 
the other high ions at positive velocities where low ionization and molecular
gas is observed.  
Much of the interstellar high ion absorption toward 3C\,273 probably
occurs within a highly fragmented medium within the Loop~IV remnant or in the 
outer cavity walls of the remnant.  Multiple ionization mechanisms are
required.

\noindent 
7) We place a flux limit of $5.4\times10^{-16}$ erg\,cm$^{-2}$\,s$^{-1}$
on the amount of \ion{O}{6} emission present in the LiF1 MDRS aperture.  
Assuming that the \ion{O}{6} emission and absorption along the sight line 
occur within the same gas, we find $n_e \lesssim 0.02$ cm$^{-3}$ and a 
thermal pressure P/k $\lesssim 11,500$ cm$^{-3}$~K for a temperature 
of $3\times10^5$~K. 

\noindent
8) An additional high velocity wing of \ion{O}{6} absorption containing
about 13\% of the total \ion{O}{6} is present 
at velocities between $+100$ and $+240$ \kms.  This absorption
wing is not observed in other species, but similar \ion{O}{6}
wings have been observed along other sight lines through the Galactic halo.
 The broad \ion{O}{6}
absorption wing may trace the expulsion of hot gas out of the Galactic disk 
into the halo.

\noindent
9) The neutral and low ionization ISM along the sight line occurs 
within two groups of clouds centered on LSR velocities of $\sim-15$ and 
$\sim+23$ \kms.  The negative velocity group contains most of the \ion{H}{1}
gas. Abundance estimates for the interstellar clouds are similar to those for 
other halo clouds.  The warm neutral and warm ionized clouds along the sight 
line have similar dust-phase abundances, implying that the properties of the 
dust grains in the two types of clouds are similar.

\noindent
10) Interstellar H$_2$ absorption is present along the 3C\,273 sight line 
at a level of $\log N$(H$_2$) $\sim 15.71$.    The H$_2$ is associated with the neutral and low ionization gas at positive
  velocities.  The H$_2$ does not trace the 
main column density concentration along the sight line observed in 
\ion{H}{1} 21\,cm emission.

\smallskip
This work was based on data obtained for the $FUSE$ Science Team by the
NASA-CNES-CSA $FUSE$ mission operated by the Johns Hopkins University.
We thank the members of the $FUSE$ operations and science teams for their 
dedicated efforts to develop and operate this wonderful observatory.
Financial support was provided by NASA contract NAS5-32985.
Partial funding for this work was also provided through NASA
Long Term Space Astrophysics grants NAG5-3485 (KRS, JCH) and NAG5-7262 (JMS).

\clearpage
\newpage
\begin{figure}[ht!]
\includegraphics{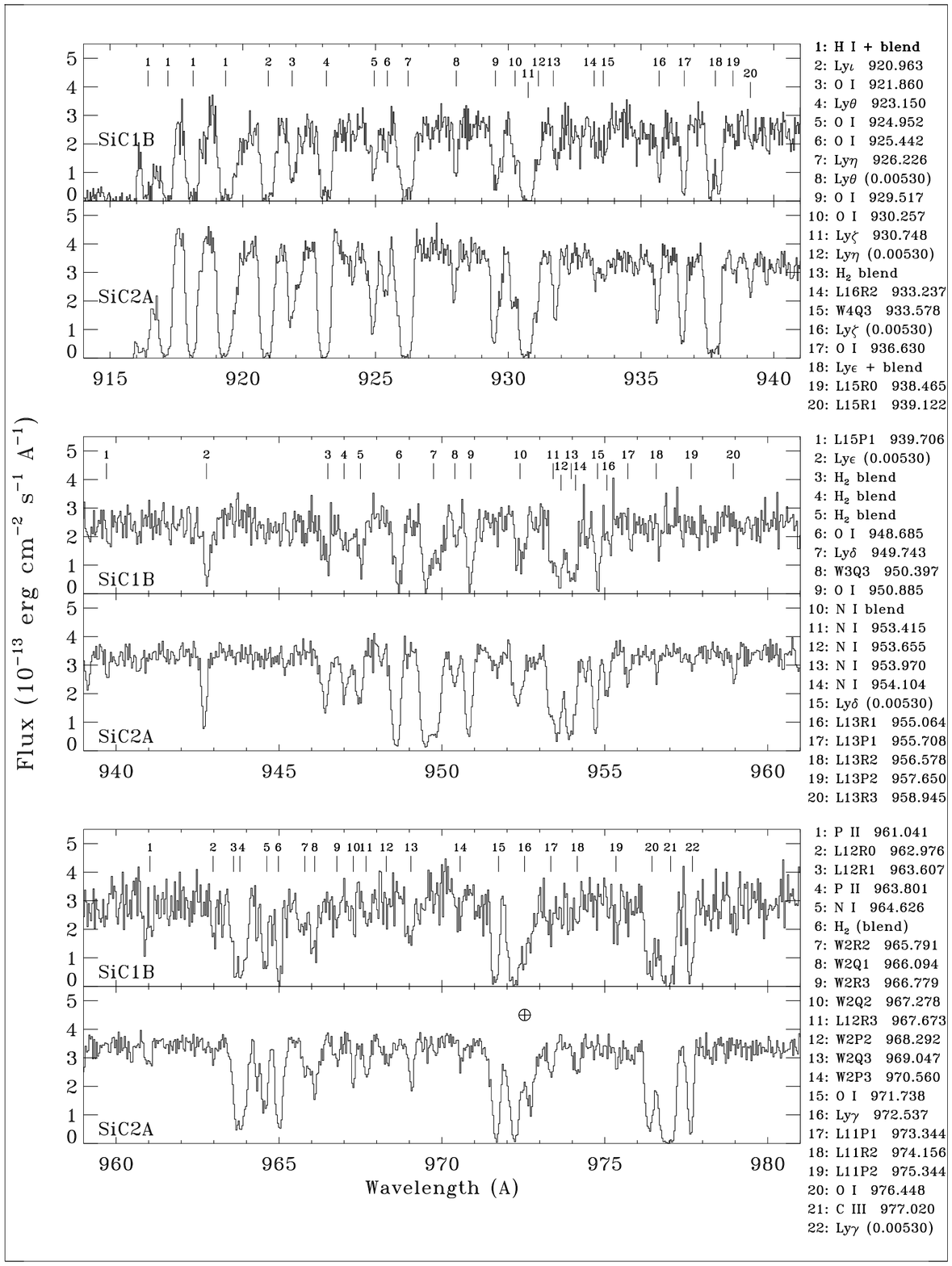}
\vspace{8.8in}
\caption{See caption at end of figure.}
\end{figure}

\clearpage
\newpage
\begin{figure}[ht!]
\figurenum{1}
\includegraphics{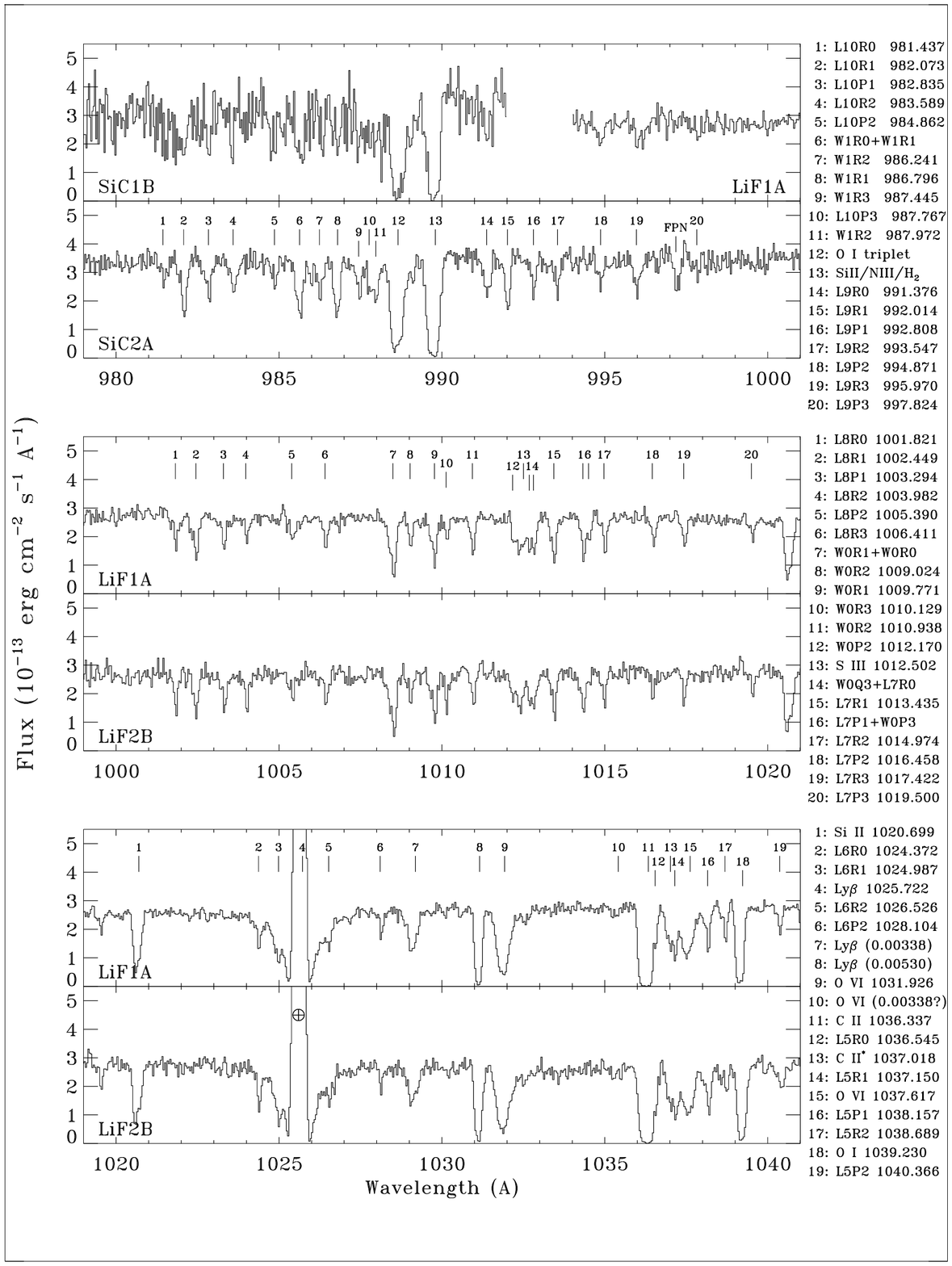}
\vspace{8.8in}
\caption{(continued)}
\end{figure}

\clearpage
\newpage
\begin{figure}[ht!]
\figurenum{1}
\includegraphics{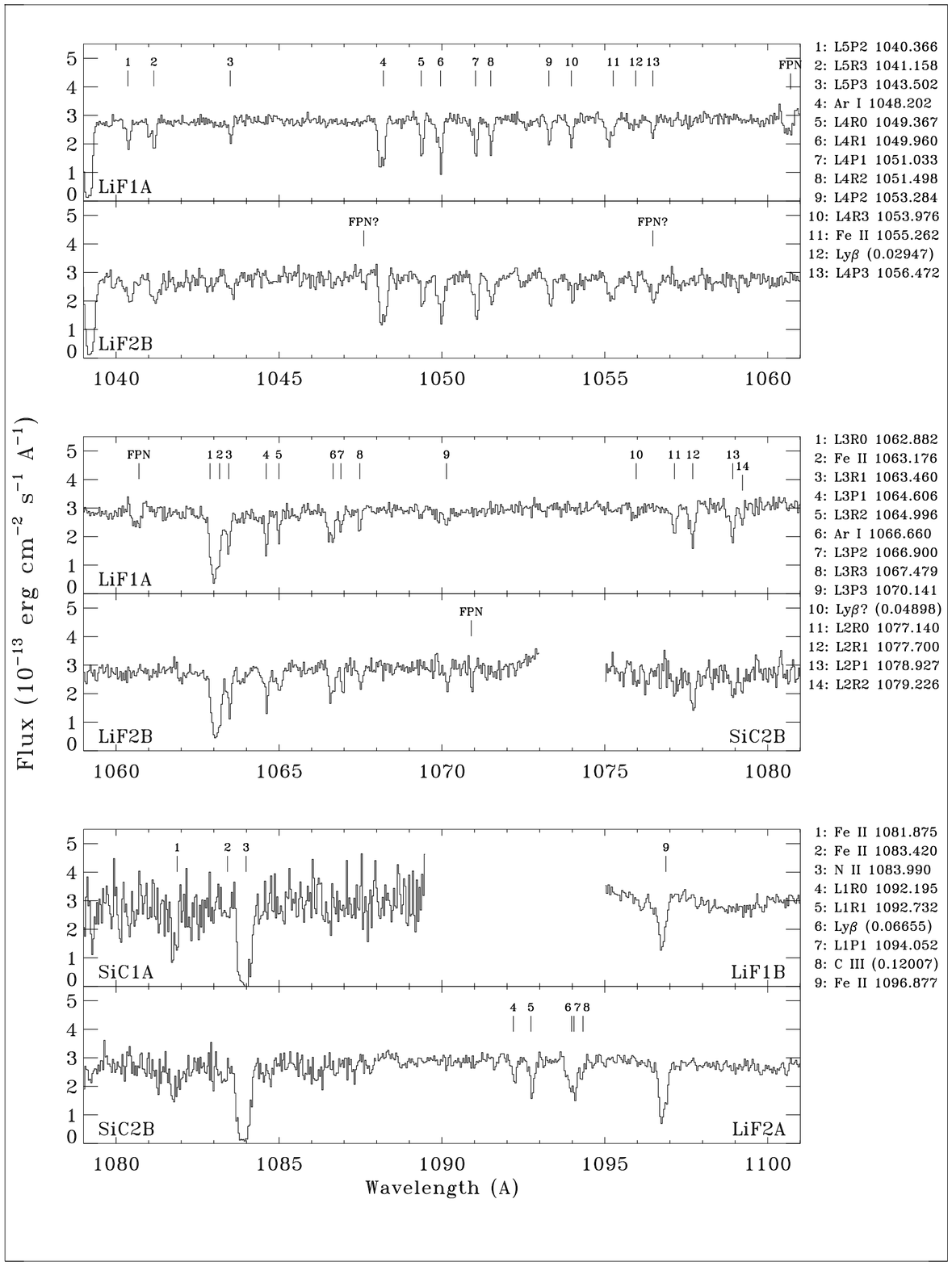}
\vspace{8.8in}
\caption{(continued)}
\end{figure}

\clearpage
\newpage
\begin{figure}[ht!]
\figurenum{1}
\includegraphics{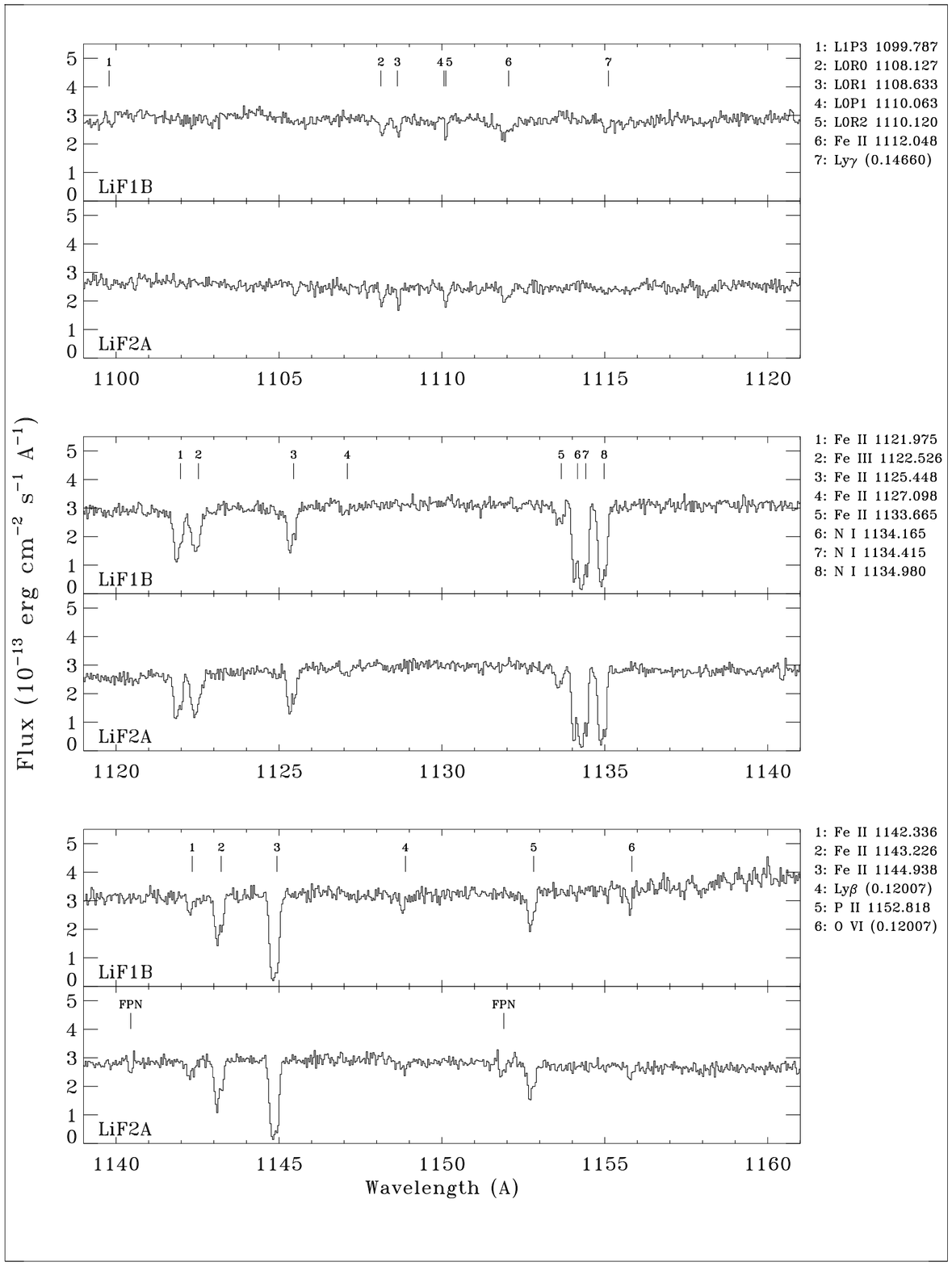}
\vspace{8.8in}
\caption{(continued)}
\end{figure}

\clearpage
\newpage
\begin{figure}[ht!]
\figurenum{1}
\includegraphics{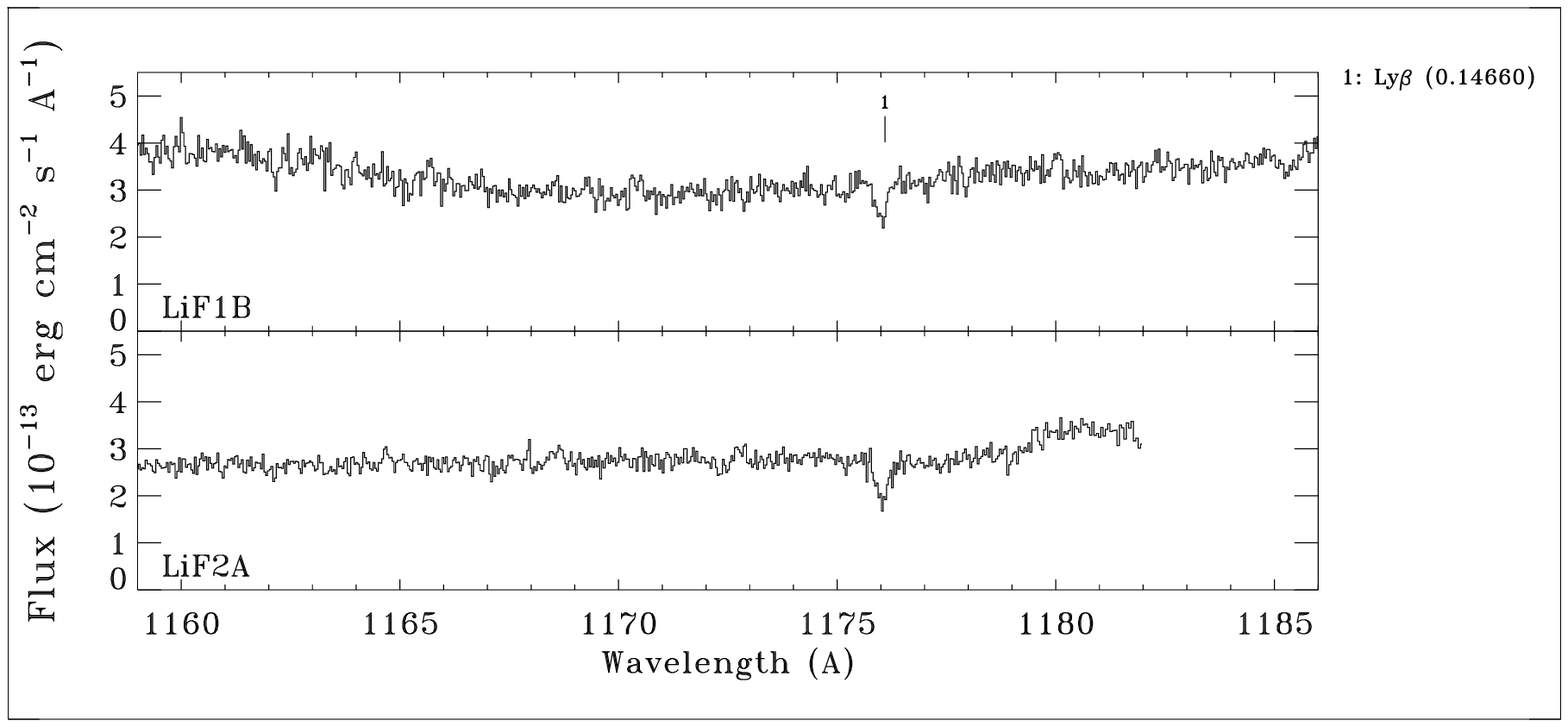}
\vspace{3.0in}
\caption{$FUSE$ spectra of 3C\,273 from 912 to 1186\,\AA.  The 
data from the 
most sensitive detector segments are shown at each wavelength 
(detector \#1 on top, detector \#2 on 
bottom).  The data have a resolution of approximately 17--22 \kms\ (FWHM), 
depending upon wavelength and detector segment.   The spectra shown have been 
binned into 0.038\,\AA\ samples and have $S/N$ $\approx$ 20--35 per resolution
element.  (This binning is larger than the binning used in the analyses
described in the text).  Prominent 
interstellar and intergalactic lines are identified to the right of each 
panel.  Molecular lines are identified by their band (Werner or Lyman), 
upper vibrational level (1-16), and rotational transition 
(R, P, or Q with lower rotational state $J=1-3$).
Numerous additional weak lines are present but are not indicated for 
clarity.
Several features resulting from fixed-pattern noise introduced by the $FUSE$
detectors are indicated with ``FPN'' above the spectra.}
\end{figure}

\clearpage
\newpage
\begin{figure}[ht!]
\figurenum{2}
\includegraphics{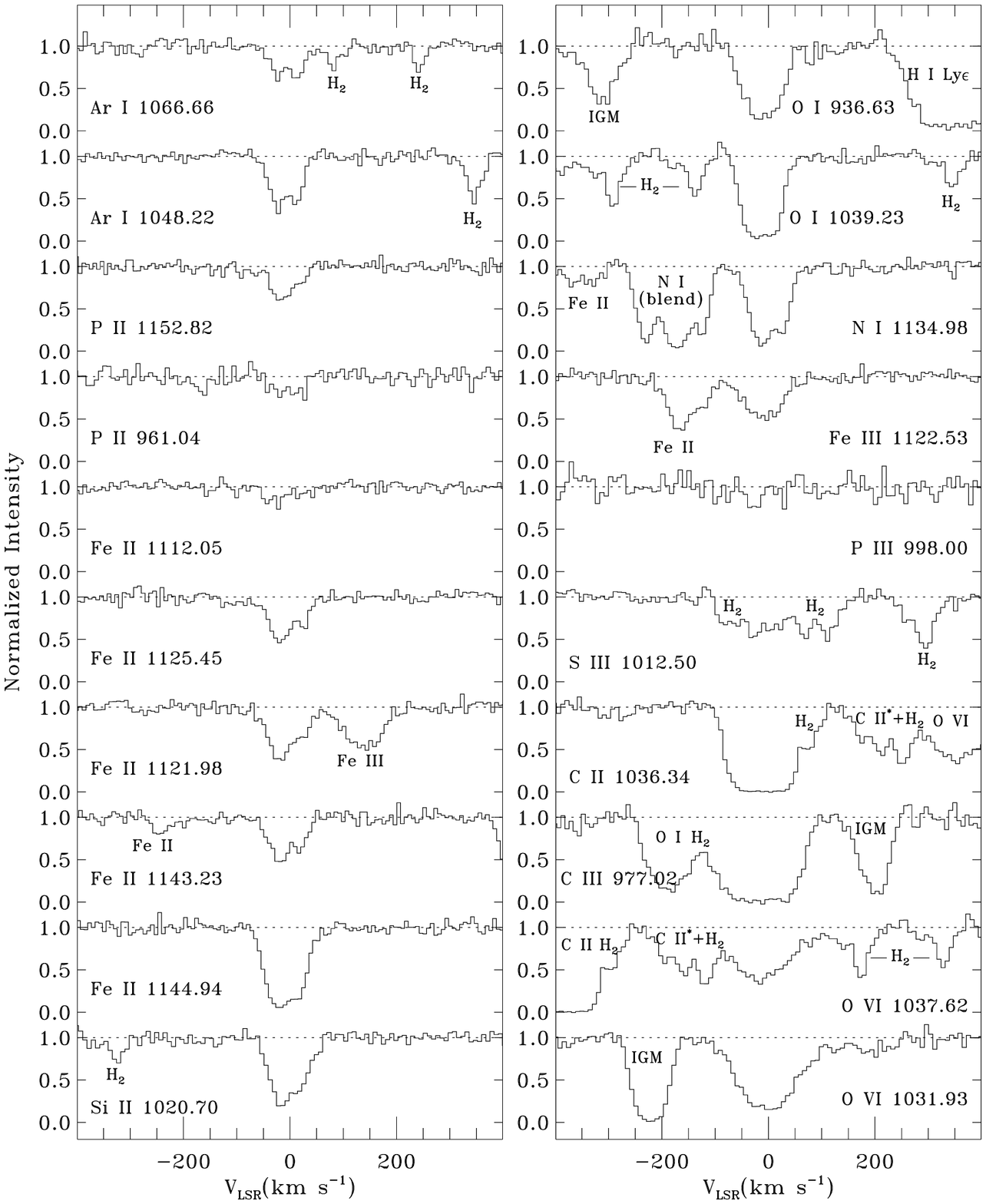}
\vspace{8.8in}
\caption{(See caption on next page.)}
\end{figure}

\clearpage
\newpage
\noindent
Figure 2.-- Continuum-normalized interstellar absorption profiles observed
by $FUSE$ along the the 3C\,273 sight line (single channel data).  The 
primary absorption features are indicated by ion and
wavelength below each spectrum.  Additional features appearing in the 
velocity range shown are labeled immediately above or below the primary 
absorption 
feature.  In a few cases, redshifted Ly$\beta$ absorption lines of intervening 
IGM clouds are present.
Note the broad shallow wing on the \ion{O}{6} $\lambda1031.926$ 
line at velocities $+100 \le v_{LSR} \le +240$ \kms. 

\clearpage
\newpage
\begin{figure}[ht!]
\figurenum{3}
\includegraphics{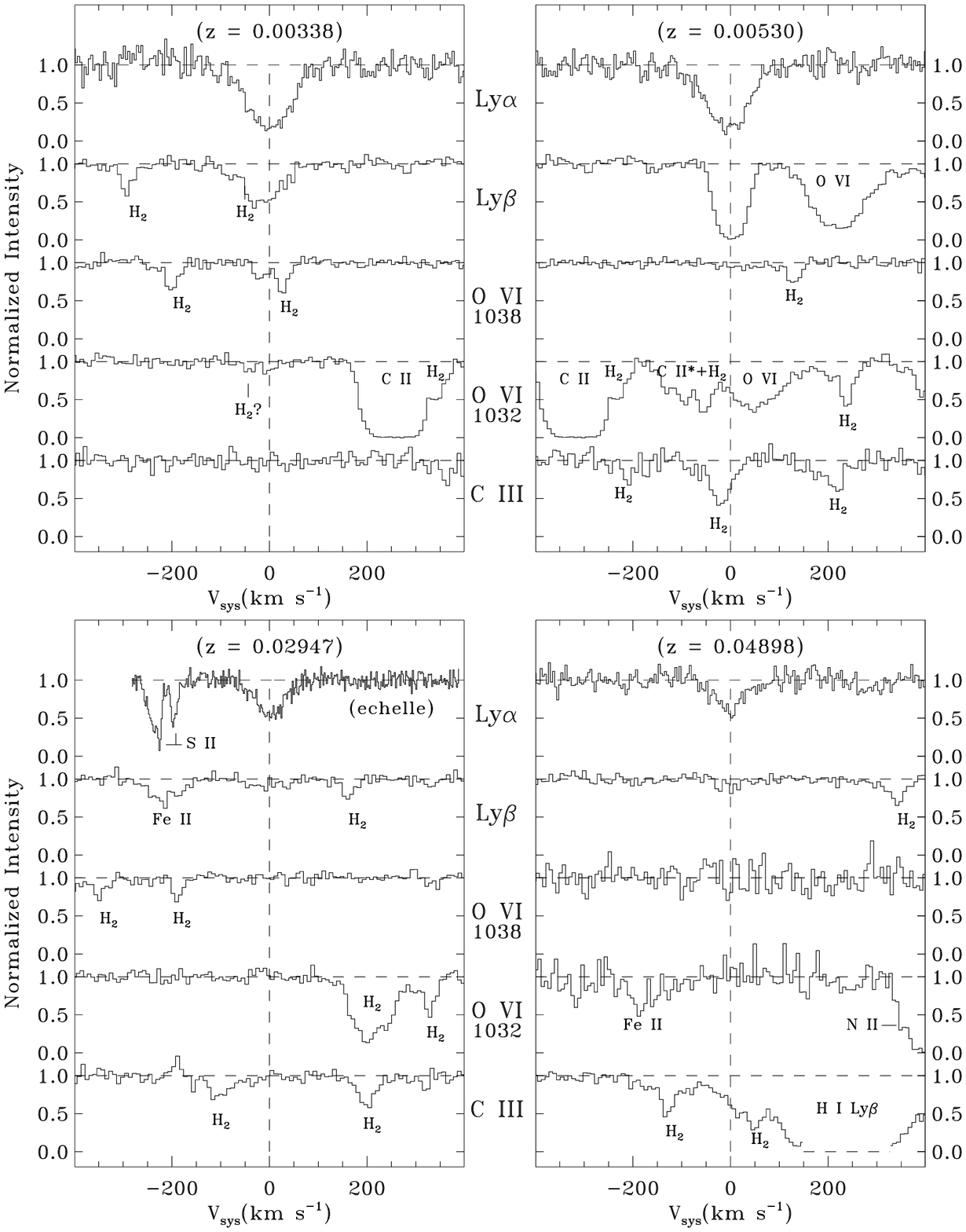}
\vspace{8.8in}
\caption{See caption at end of figure.}
\end{figure}

\clearpage
\newpage
\begin{figure}[ht!]
\figurenum{3}
\includegraphics{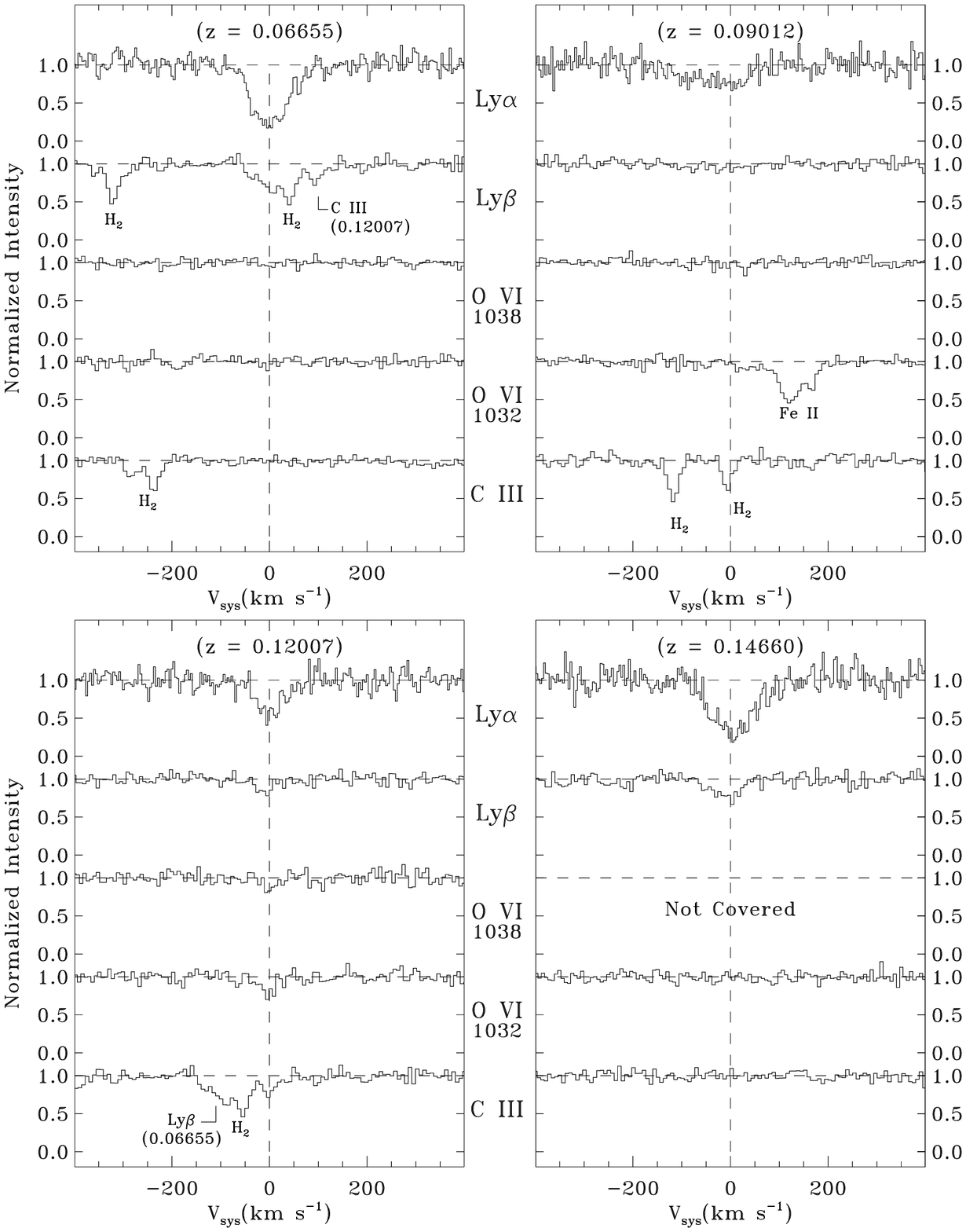}
\vspace{8.8in}
\caption{(continued)}
\end{figure}

\clearpage
\newpage
\noindent
Fig. 3.-- Continuum-normalized Ly$\alpha$, Ly$\beta$, \ion{O}{6}
$\lambda1037.617$, \ion{O}{6} $\lambda1031.926$, and 
\ion{C}{3} $\lambda977.020$ 
absorption profiles for eight of the previously identified Ly$\alpha$
absorbers along the 3C\,273 sight line as a function of rest frame velocity.
Only data from one channel is shown for each line observed by $FUSE$.
The redshift of the absorption is indicated at the top of each panel.
The Ly$\alpha$ 
profiles are pre-COSTAR GHRS data except for the $z=0.02947$ system, for
which GHRS echelle (FWHM $\approx$ 3.5 \kms) data were available.  Ly$\beta$
absorption is detected for each system.  \ion{O}{6} 
absorption is detected in the $z=0.00338$ and $z=0.12007$ absorbers.  
\ion{C}{3} absorption is detected only in the $z=0.12007$ absorber.  
The \ion{O}{6} $\lambda1037.617$ line in the $z=0.14660$ absorber 
was not 
covered by the $FUSE$ bandpass.  Galactic lines are identified immediately 
above or below each spectrum.

\clearpage
\newpage
\begin{figure}[ht!]
\figurenum{4}
\includegraphics{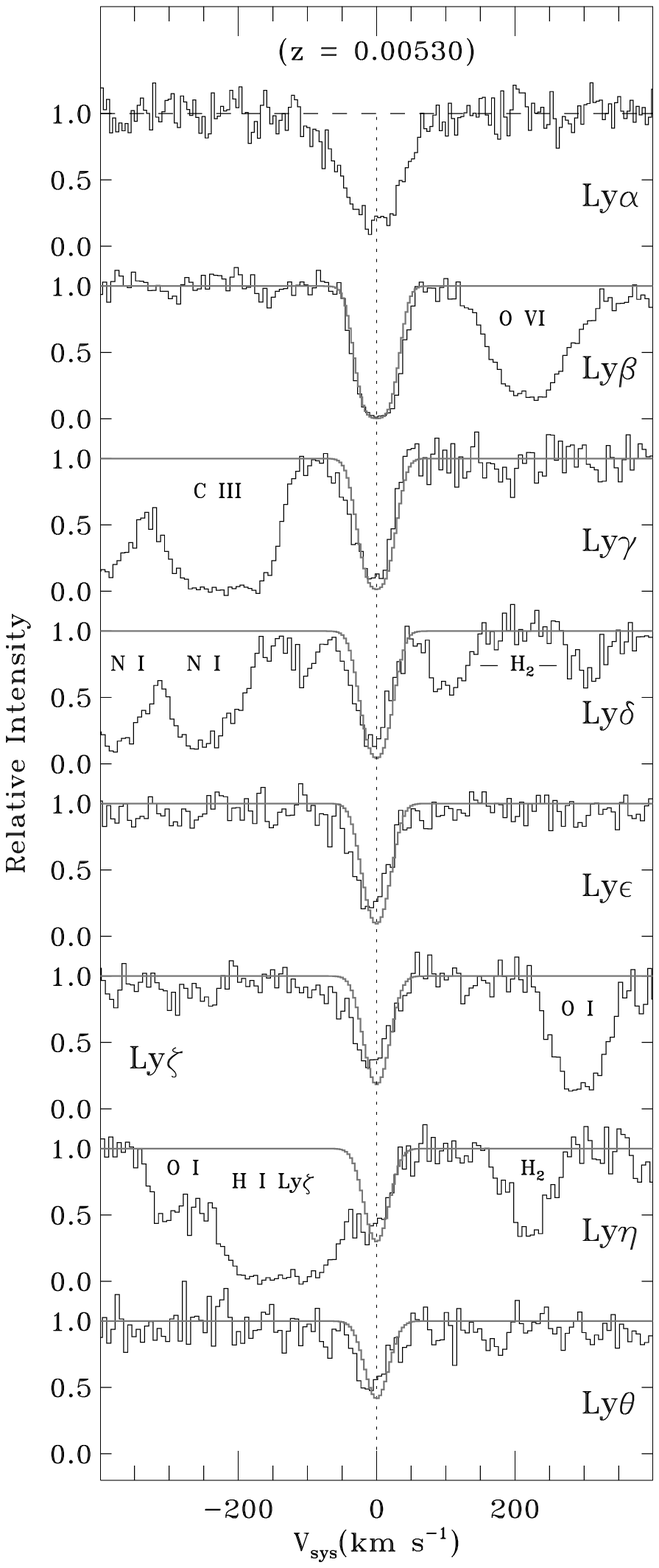}
\vspace{8.5in}
\caption{Caption on next page.} 
\end{figure}

\clearpage
\newpage
\noindent
Fig. 4.--Continuum-normalized \ion{H}{1} Lyman series absorption profiles 
for the $z = 0.00530$ (1590 \kms) Virgo absorber as observed by $FUSE$.  
The Ly$\alpha$ profile shown at the top of the plot is a pre-COSTAR GHRS 
spectrum.  Note that the $FUSE$ data are of higher resolution than the 
GHRS data since the GHRS data were obtained with an instrumental line spread
function having very broad and strong 
($\sim40$ \kms) wings.  The Ly$\beta$ absorption 
in this system is very strong compared to the other absorbers along the 
sight line.  The light grey lines overplotted on the spectra are the 
profiles constructed with the best fit single-component curve of 
growth parameters listed in Table~4 (see \S4.1).  These synthetic spectra
have been convolved with a Gaussian instrumental function with
$FWHM=0.075$\,\AA.  The actual line spread function may vary slightly 
with $\lambda$.  Galactic lines are 
identified immediately above or below each spectrum.

\clearpage
\newpage
\begin{figure}[ht!]
\figurenum{5}
\includegraphics{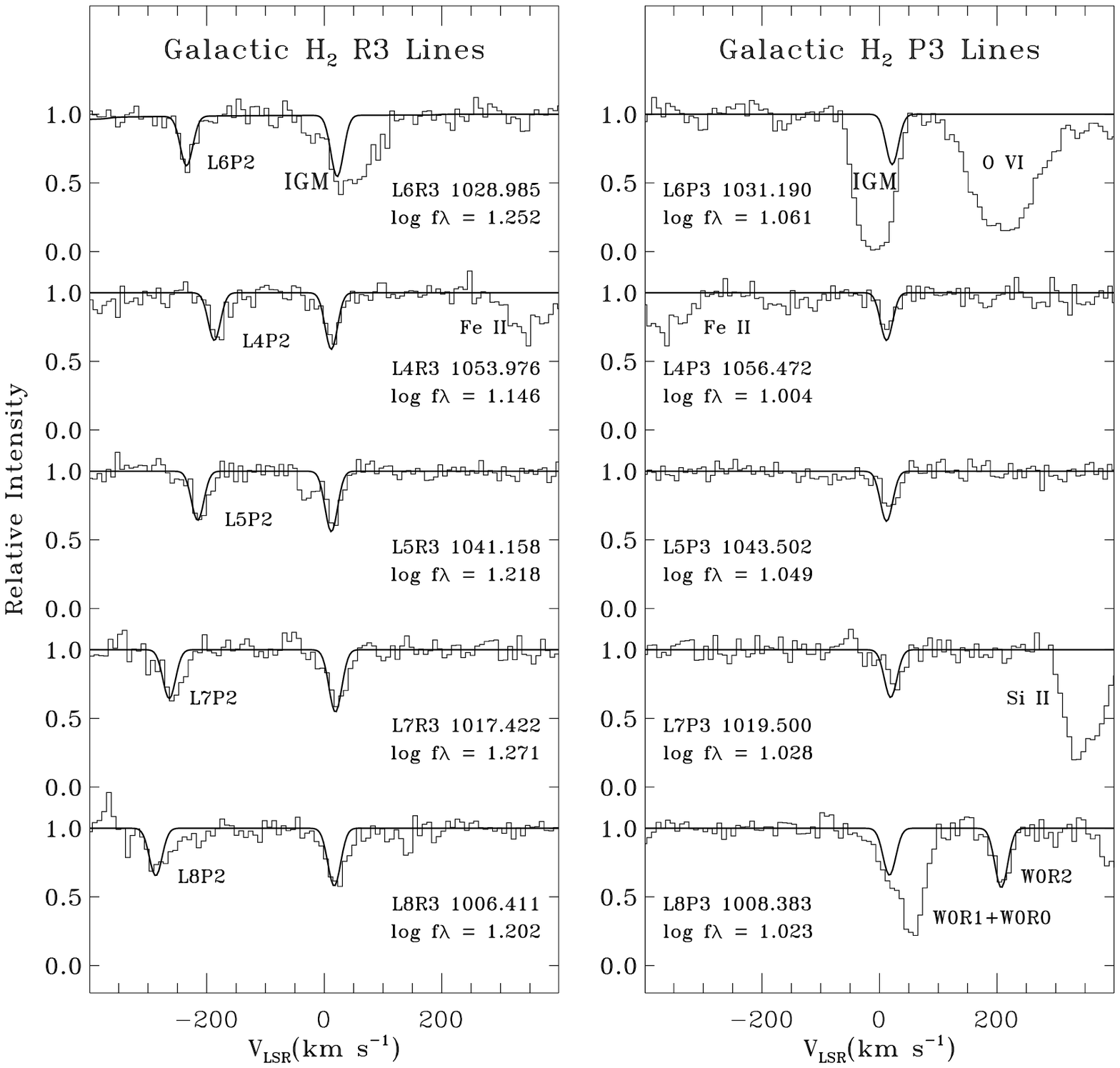}
\vspace{6.5in}
\caption{A comparison of the H$_2$ (6--0) 
R(3) and P(3) lines in the vicinity 
of the Virgo IGM absorbers with H$_2$ $J=3$ lines of similar strength in other 
regions of the $FUSE$ spectra of 3C\,273.  The lines are plotted as a function
of LSR velocity of the H$_2$ lines.  For each line, the wavelength and 
product $\log f\lambda$ are listed.  Additional H$_2$ and atomic lines 
covered in the plots are also labeled.  The strong absorption features 
at the top of each panel are the $z=0.00338$ Virgo absorber (left panel)
and the $z=0.00530$ Virgo absorber (right panel).}
\end{figure}

\clearpage
\newpage
\begin{figure}[ht!]
\figurenum{6}
\includegraphics{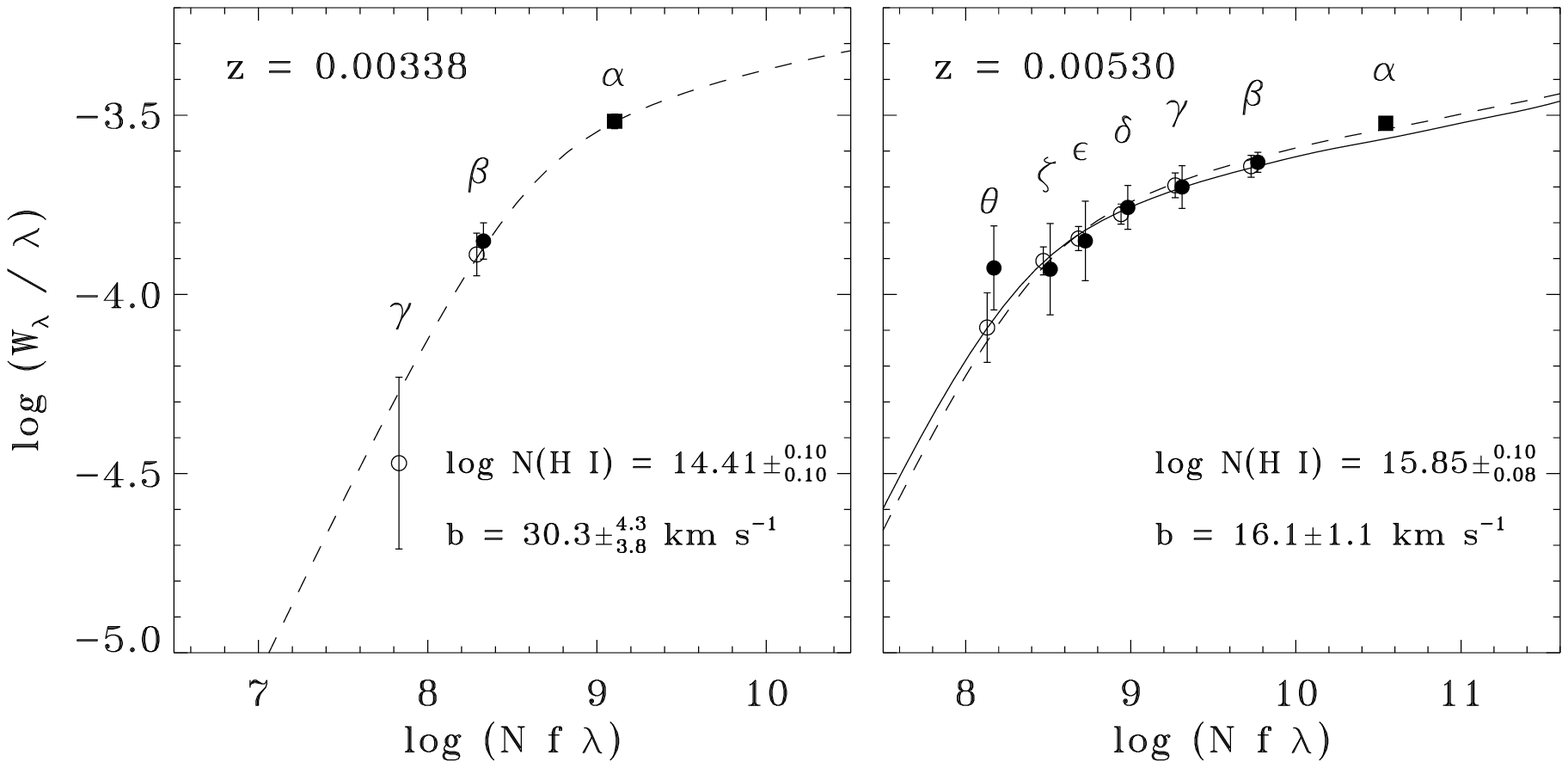}
\vspace{5.0in}
\caption{Single-component, Doppler-broadened curves of growth fitted to the 
\protect\ion{H}{1} lines observed in the $z = 0.00338$ (1015 \kms) and $z=0.00530$
(1590 \kms) Virgo Cluster absorbers.  The data points include 
GHRS data for Ly$\alpha$ (solid squares) and $FUSE$ data for \protect\ion{H}{1}
lines on detectors 1 and 2 (filled and open circles, respectively).  
For the $z=0.00338$ absorber, the fit shown includes the Ly$\alpha$
data point in the calculation.  For the $z=0.00530$ absorber, two fits
are shown -- one based solely on the $FUSE$ data (solid line), and one
including the Ly$\alpha$ data (dashed line). 
The best fit column density for the $z=0.00533$ absorber based on the $FUSE$ 
data is a factor of 43 
higher than previous estimates based on Ly$\alpha$ profile fitting alone.
}
\end{figure}

\clearpage
\newpage
\begin{figure}[ht!]
\figurenum{7}
\includegraphics{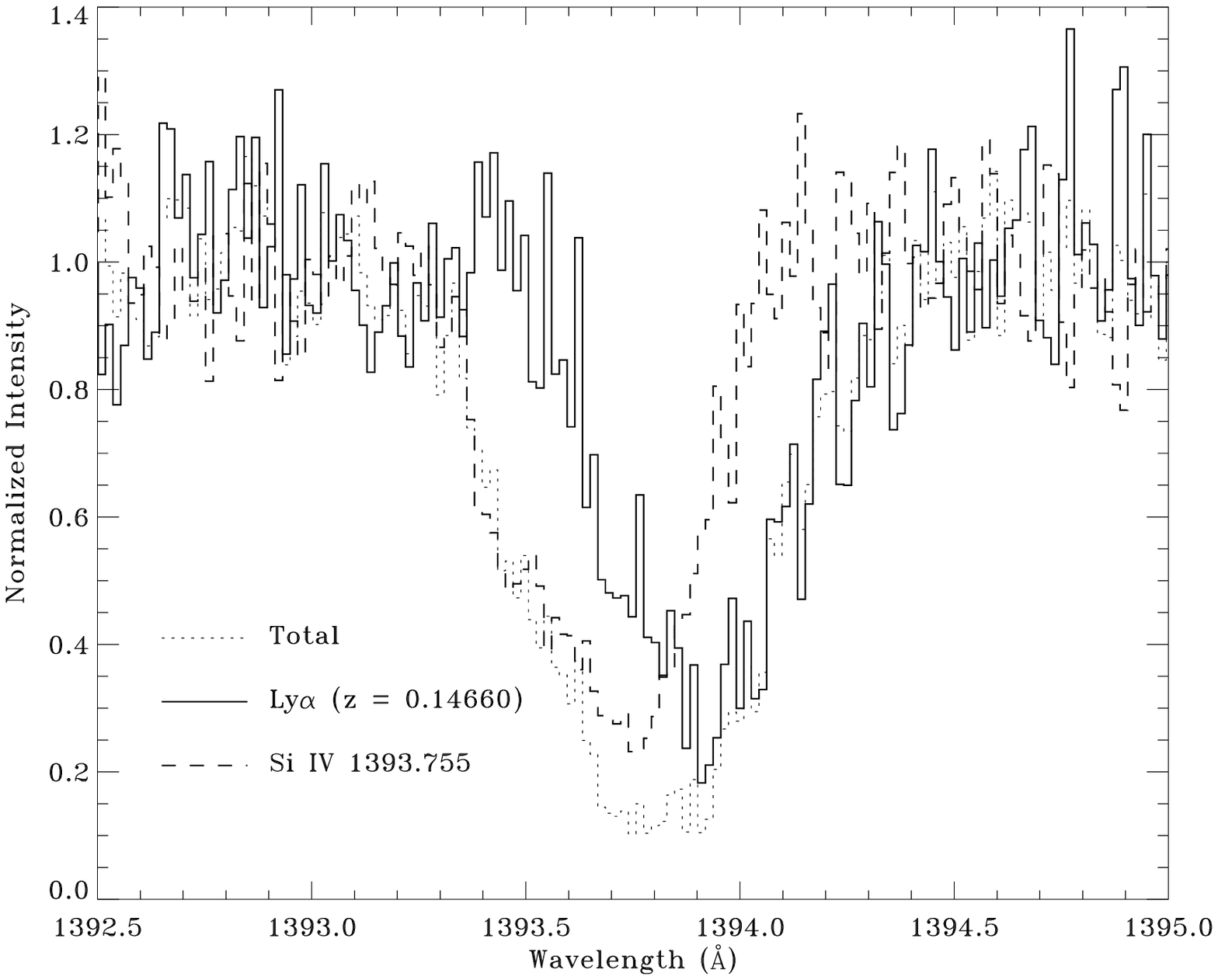}
\vspace{6.0in}
\caption{Decomposition of the $z=0.14660$ Ly$\alpha$ absorber and the 
Galactic \protect\ion{Si}{4} $\lambda1393.755$ line plotted as a function of 
LSR velocity.  The dotted profile is the observed intergalactic 
Ly$\alpha$ plus interstellar \protect\ion{Si}{4} $\lambda1393.755$ absorption.
The solid and dashed lines show the decomposition into the IGM and ISM
components.  This decomposition differs from the one shown by 
Savage et al. (1993) - see \S4.7 for details.}
\end{figure}

\clearpage
\newpage
\begin{figure}[ht!]
\figurenum{8}
\includegraphics{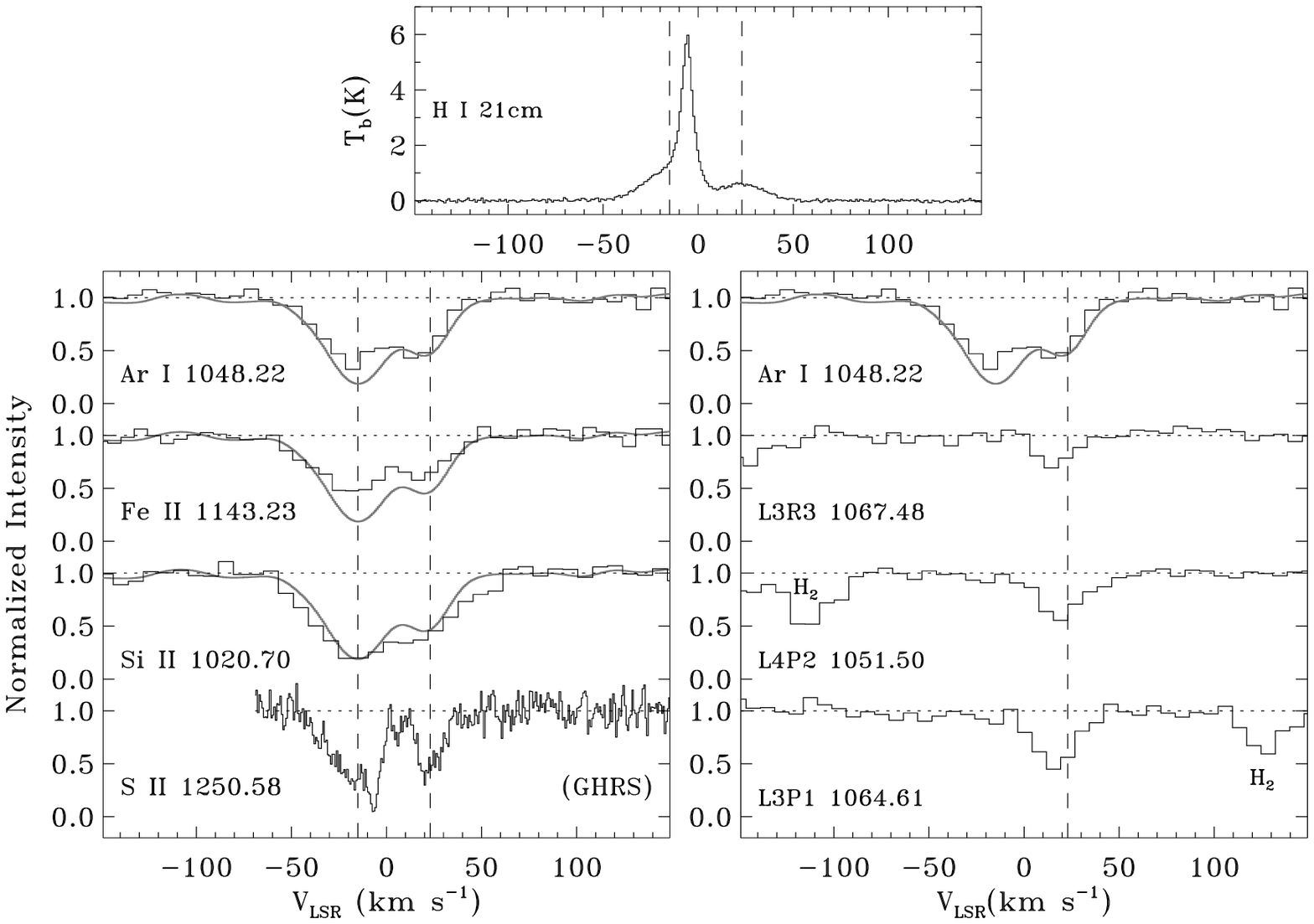}
\vspace{6.0in}
\caption{Low ion comparison for selected interstellar atomic and H$_2$ 
lines shown in Figures~1 and 2. 
The vertical dashed lines indicate the centroids of the negative and 
positive velocity groupings of clouds along the sight line.  The light
curve shown on top of the \protect\ion{Ar}{1}, \protect\ion{Fe}{2}, and 
\protect\ion{S}{2} profiles 
is the high-resolution (GHRS Ech-A)
\protect\ion{S}{2} $\lambda1253.811$ line smoothed to a $FUSE$ resolution of 
$\sim20$ \kms.  Note that the H$_2$ absorption is closely aligned in velocity 
with the weaker of the two cloud groupings.}
\end{figure}

\clearpage
\newpage
\begin{figure}[ht!]
\figurenum{9}
\includegraphics{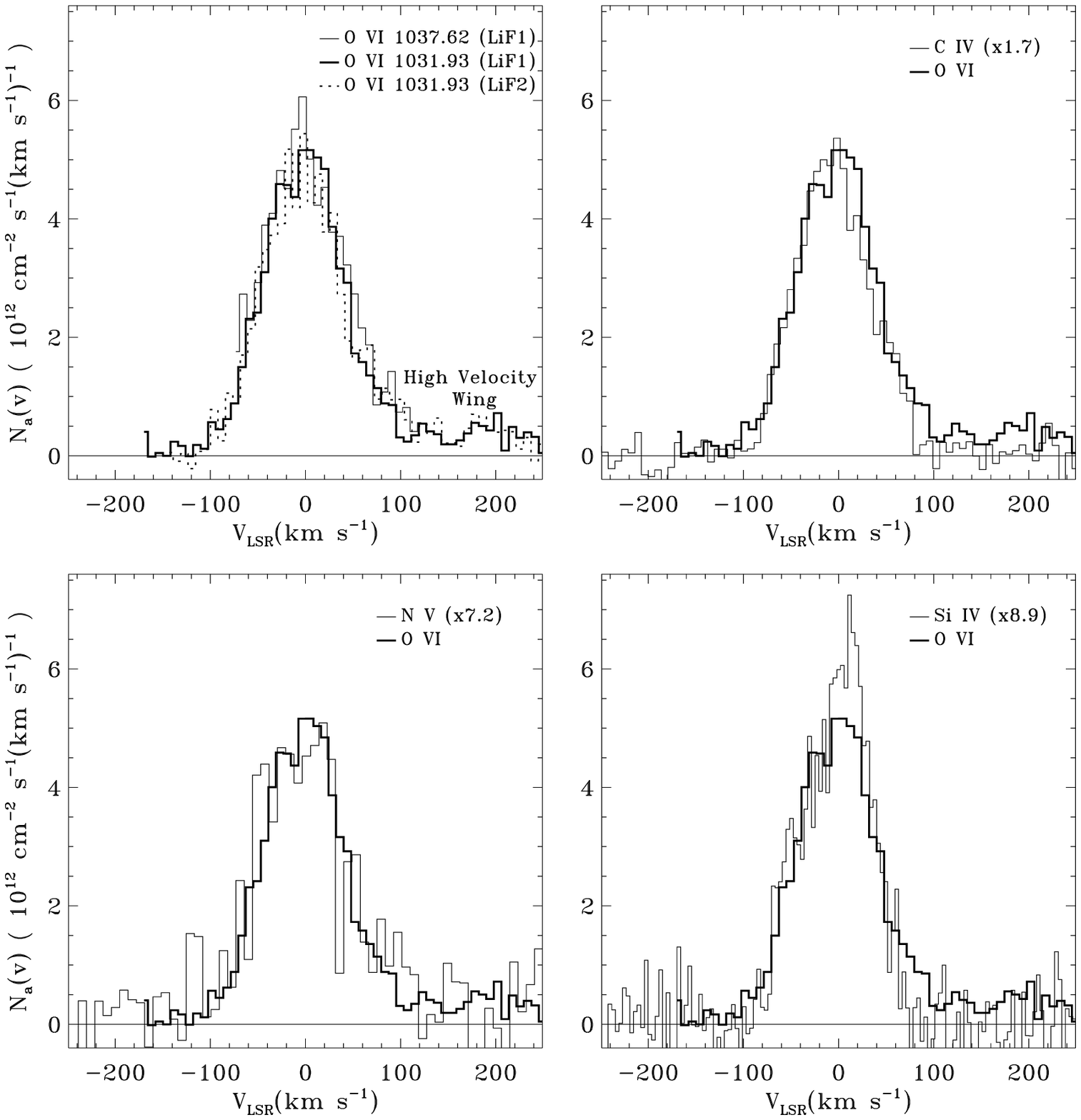}
\vspace{6.8in}
\caption{{\it Top left panel:} Apparent column density profiles versus LSR
velocity for the interstellar \protect\ion{O}{6} $\lambda\lambda1031.926, 
1037.617$ 
lines observed by $FUSE$.  The two lines differ in the quantity $\log f\lambda$
by a factor of 2.  The excellent agreement of the profiles indicates
that there is no unresolved saturated structure within the line profiles.
{\it Remaining panels:} Comparison of the \protect\ion{O}{6} $\lambda1031.926$ 
apparent
column density profile with N$_a(v)$ profiles derived from GHRS data for 
\protect\ion{C}{4} $\lambda1550.770$ ({\it top right}), 
\protect\ion{N}{5} $\lambda1238.821$
({\it bottom left}),  and \protect\ion{Si}{4} $\lambda1402.770$ 
({\it bottom right}).  
The \protect\ion{C}{4}, \protect\ion{N}{5},  and \protect\ion{Si}{4}
profiles have been scaled by factors equal to the values given in Eq. (5).}
\end{figure}

\clearpage
\newpage
\begin{figure}[ht!]
\figurenum{10}
\includegraphics{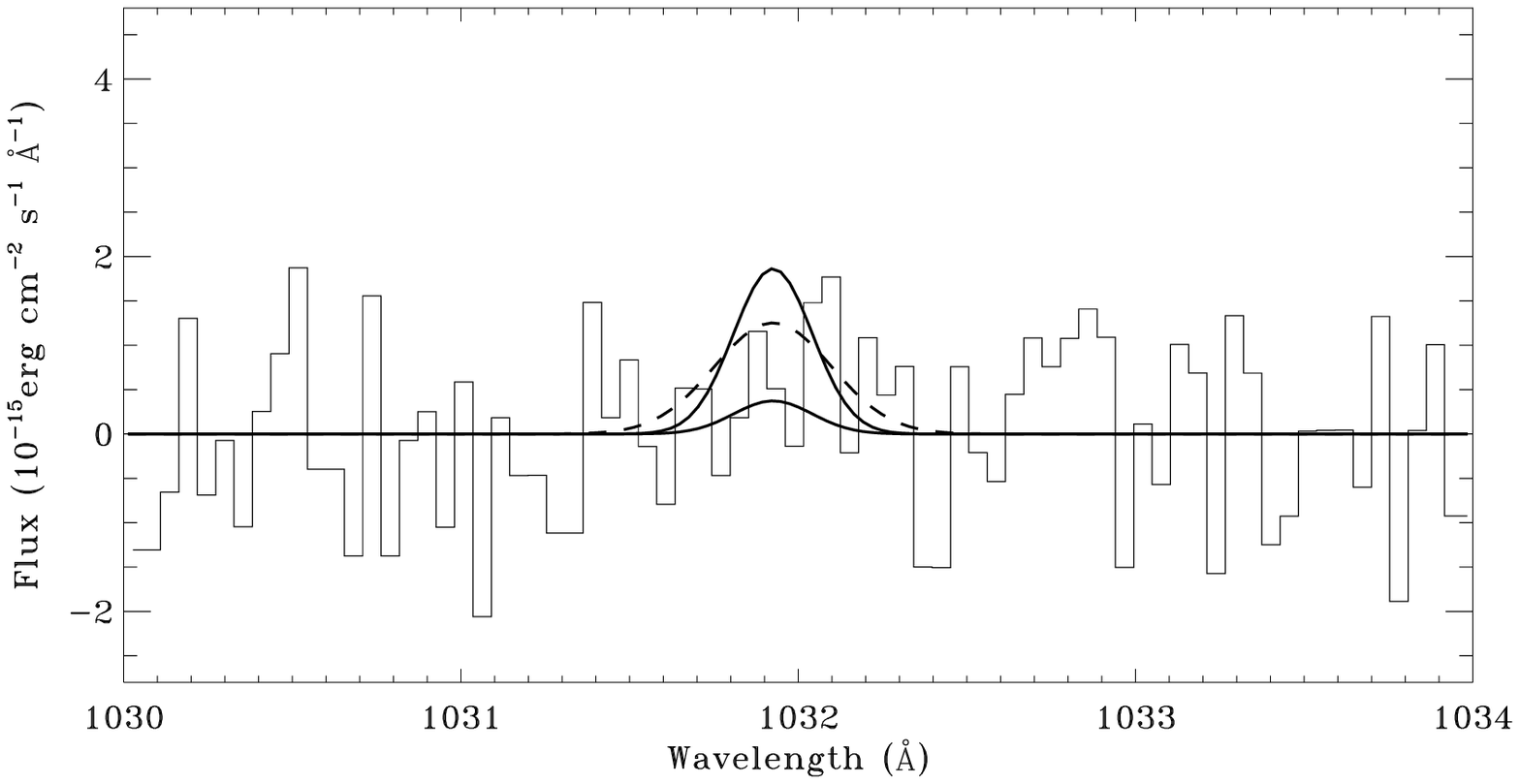}
\vspace{4.0in}
\caption{Extracted sky spectrum toward 3C\,273 from the LiF1 MDRS 
aperture.  The data (histogrammed line) have been binned into $\sim0.06$\,\AA\ 
bins, and a  constant ``background'' of $5\times10^{-16}$ 
erg\,cm$^{-2}$\,s$^{-1}$\,\AA$^{-1}$ has been subtracted from the nominally
calibrated pipeline flux.  The three smooth 
curves shown have the following parameters --  solid lower profile:
I$_{\lambda1032}$ = 3000 ph\,cm$^{-2}$\,s$^{-1}$\,sr$^{-1}$, FWHM = 80 \kms;
solid upper profile: I$_{\lambda1032}$ = 
15,000 ph\,cm$^{-2}$\,s$^{-1}$\,sr$^{-1}$, FWHM = 80 \kms; dashed profile:
I$_{\lambda1032}$ = 
15,000 ph\,cm$^{-2}$\,s$^{-1}$\,sr$^{-1}$, FWHM = 120 \kms.}

\end{figure}

%
%
 
\begin{deluxetable}{lcccccc}
\tablecolumns{7}
\tablewidth{375pt} 
\tablecaption{Selected Interstellar Atomic Absorption Lines Toward 3C\,273\tablenotemark{a}}
\tablehead{Species & $\lambda$ & $\log f\lambda$  & W$_\lambda$(Det1) & W$_\lambda$(Det2) & Vel. Range & Note\tablenotemark{b}\\
& (\AA) && (m\AA) & (m\AA) & (\kms)}
\startdata
\ion{H}{1}  & 1025.722 & 1.909 & $\sim1540$ & $\sim1540$ & \nodata & 1 \\
\\
\ion{C}{2}  & 1036.337 & 2.106 & $474\pm15$ & $479\pm17$ &--150 to +40 & 2\\
\\
\ion{C}{3}  & \phn977.020 & 2.872 & $533\pm34$ & $562\pm13$ & --140 to +100\\
\\
\ion{N}{1}  & \phn964.626 & 1.002 & $157\pm24$ & $157\pm10$ & --80 to +30 \\
	    & 1134.980 & 1.660 & $276\pm10$ & $259\pm10$ &--100 to +70 \\
\\
\ion{O}{1}  & 1039.230 & 0.980 & $283\pm10$ & $314\pm13$ &--100 to +80 \\
\\
\ion{O}{6}  & 1037.617 & 1.836 & $>240$ & $>240$ & \nodata & 3 \\
	    & 1031.926 & 2.137 & $393\pm15$ &  $392\pm16$ & --160 to +100 & 4 \\
	    & 	       &       & $\phn57\pm10$ & $\phn57\pm12$ & +100 to +240 & 34\\
\\
\ion{Si}{2} & 1020.699 & 1.460 & $199\pm10$ & $203\pm10$ &--100 to +60 \\
\\
\ion{P}{2}  & 1152.818 & 2.435 & $\phn95\pm10$ & $\phn97\pm10$ &  --100 to +40 \\
\\
\ion{S}{2}  & 1250.584 & 0.834 & \multicolumn{2}{c}{$140\pm07$} &--60 to +40 & 5 \\
 	    & 1253.811 & 1.135 & \multicolumn{2}{c}{$197\pm05$} &--60 to +40 & 5 \\
	    & 1259.519 & 1.311 & \multicolumn{2}{c}{$235\pm14$} &--60 to +40 & 5 \\
\\
\ion{S}{3}  & 1012.502 & 1.556 & $113\pm10$ & $111\pm13$ &--65 to +30 & 6 \\
\\
\ion{S}{6}  & \phn944.523 & 2.314 & $\phn45\pm41$ & $\phn13\pm13$ & --100 to +100 & 7\\
\\
\ion{Ar}{1} & 1066.660 & 1.851 & $\phn89\pm10$ & $\phn76\pm10$ &--120 to +40 \\
	    & 1048.220 & 2.408 & $134\pm10$ & $145\pm10$ &--90 to +40 \\
\\
\tableline
\\
\\
\ion{Fe}{2} & 1127.098 & 0.583 & $\phn20\pm10$ & $\phn25\pm10$ &--100 to +40 \\
	    & 1133.665 & 0.795 & $\phn56\pm10$ & $\phn54\pm10$ & --100 to +40 \\
	    & 1055.262 & 0.898 & $\phn75\pm10$ & $\phn69\pm10$&--100 to +40\\
	    & 1112.048 & 1.006 & $\phn73\pm10$ & $\phn50\pm10$ &--100 to +40 & 8\\
	    & 1125.448 & 1.255 & $137\pm10$ & $148\pm10$&--120 to +60 & 9 \\
	    & 1143.226 & 1.308 & $131\pm10$ & $144\pm10$ &--100 to +40 \\
	    & 1121.975 & 1.329 & $165\pm10$ & $154\pm10$ &--100 to +40 \\
	    & 1096.877 & 1.545 & \nodata & $201\pm10$ &--100 to +40 & 10 \\
	    & 1144.938 & 2.084 & $302\pm10$ & $297\pm10$ &--100 to +40 \\
\\
\ion{Fe}{3} & 1122.526 & 1.947 & $160\pm10$ & $172\pm10$ &--100 to +50\\
\enddata
\tablenotetext{a}{Wavelengths and $f$-values are from Morton (1991).
Errors are $1\sigma$ estimates and limits are $3\sigma$ estimates 
unless otherwise indicated.}
\tablenotetext{b}{Notes: (1) Equivalent widths derived from a fit to the damping 
wings of the Ly$\beta$ line, for which log N(\ion{H}{1}) = 20.20.  (2) H$_2$ Lyman 
(5--0) R(0) occurs at +60 \kms\ with respect to \ion{C}{2}.
(3)Equivalent width is quoted as a limit for comparison with 
$\lambda1031.926$ since it was derived over the limited velocity range of --90 to +60 \kms.
(4) Total \ion{O}{6} $\lambda$1031.926 equivalent width (--160 to +240 \kms) is 450$\pm$15 m\AA.
(5) \ion{S}{2} values from HST data.  The $\lambda1250.584$ and $\lambda1253.811$ are 
from post-COSTAR GHRS Ech-A data, and the $\lambda1259.519$ value is from pre-COSTAR GHRS G160M data.  
(6) H$_2$ Werner (0--0) P(2) at 1012.17\,\AA\ (--99 \kms\ w.r.t. \ion{S}{3}),
W0Q3 at 1012.68\,\AA, and L7R0 at 1012.81\,\AA\ bracket the \ion{S}{3}
line.
(7) Stronger \ion{S}{6} line at 933.378 \AA\ is blended with 
H$_2$ L16R2 and W4Q3 lines.
(8) Line appears as a much broader feature in the LiF1B 
spectrum.  The equivalent widths
of this feature are uncertain since the feature should be at least as strong 
as the $\lambda1055.262$ 
line.
(9) Feature centered at --122 \kms\ with W$_\lambda$ $\approx$ 
15 m\AA\ appears in both
channels and is of unknown origin.  It is not likely to be \ion{Fe}{2} 
because it is not present in the stronger \ion{Fe}{2} lines.
(10) Line falls near edge of segment LiF1B, leading to an 
unreliable value of W$_\lambda$ on detector 1.}
\end{deluxetable}

\clearpage
\newpage
\begin{deluxetable}{cccccc}
\tablecolumns{6}
\tablewidth{0pt} 
\tablecaption{Selected Interstellar H$_2$ Absorption Lines Toward 3C\,273}
\tablehead{Transition\tablenotemark{a}  & $\lambda$ & $\log f\lambda$  
& W$_\lambda$(Det1)\tablenotemark{b} & W$_\lambda$(Det2)\tablenotemark{b} & log\,N \\
& (\AA) & & (m\AA) & (m\AA) & (3$\sigma$)}
\startdata
\multicolumn{6}{c}{$J=3$} \\
\tableline
L3R3 & 1067.480 & 1.033 & 21$\pm$5 & 37$\pm$9 & \nodata\\
L4R3 & 1053.980 & 1.146 & 45$\pm$5 & 42$\pm$9 & \nodata \\
L5R3 & 1041.160 & 1.218 & 57$\pm$6 & $<74$\tablenotemark{c} & \nodata \\
L7R3 & 1017.420 & 1.271 & 56$\pm$6 & 44$\pm$8 & \nodata \\
L7P3 & 1019.500 & 1.028 & 35$\pm$7 & 33$\pm8$ & \nodata \\
L8R3 & 1006.410 & 1.202 & 64$\pm$15 & 55$\pm$13  & \nodata\\
W0R3 & 1010.130 & 1.143 & 36$\pm$7 & 33$\pm$8 & \nodata \\
L0P3 & 1099.790 & 0.432 & 12$\pm$6 & 14$\pm$5 & \nodata \\
L3P3 & 1070.140 & 0.903 & 32$\pm$6 & 36$\pm$7 & \nodata \\
L4P3 & 1056.470 & 1.004 & 30$\pm$5 & 44$\pm$10 & \nodata \\
W2P3 & \phn970.560 & 0.978 & \nodata & 31$\pm$9 & \nodata \\
W3Q3 & \phn950.400 & 1.411 & \nodata & 50$\pm$10 & \nodata \\
\cutinhead{$J=4$}
L3R4 & 1070.900 & 1.015 & \phn6$\pm$10 & 12$\pm$12 & $<$14.49 \\
L4R4 & 1057.380 & 1.135 & 16$\pm$10    & 14$\pm$10 & $<$14.37 \\
L5P4 & 1047.550 & 1.062 & \phn3$\pm$10 & --5$\pm$12 & $<$14.44 \\
L5R4 & 1044.540 & 1.206 & \phn3$\pm$10 & --2$\pm$12 & $<$14.31\\
W0R4 & 1011.810 & 1.147 & \phn8$\pm$10 & \phn8$\pm$12 & $<$14.38 \\
L9R4 & \phn999.270 & 1.221 & \phn8$\pm$13 & \phn0$\pm$16 & $<$14.42 \\
W1P4 & \phn994.230 & 1.138 & --8$\pm$15 & 26$\pm$17 & $<$14.57 \\
\enddata
\tablenotetext{a}{Notation for the H$_2$ lines is as follows: ``L'' indicates 
Lyman series transitions. ``W'' indicates Werner series transitions.  The 
first numerical digit indicates the upper vibrational state of the
transition (the lower vibrational state is 0 in all cases).  The last two 
characters indicate the rotational transition selection rule (P, Q, R) 
and the upper rotational level (3 or 4).}
\tablenotetext{b}{Errors are $1\sigma$ estimates.}
\tablenotetext{c}{Absorption is broader and stronger than expected.  Possible 
fixed-pattern noise contamination.}
\end{deluxetable}

\clearpage
\newpage
\begin{deluxetable}{lccccl}
\tablecolumns{6}
\tablewidth{524pt} 
\tablecaption{Intergalactic Absorption Lines Toward 3C\,273\tablenotemark{a}}
\tablehead{Species & $\lambda_{rest}$ & $\lambda_{obs}$ & W$_{obs}$(\#1)\tablenotemark{b} & W$_{obs}$(\#2)\tablenotemark{b} 
& \multicolumn{1}{c}{Comments\tablenotemark{c}} \\
& (\AA) & (\AA) & (m\AA) & (m\AA)}
\startdata
\multicolumn{6}{c}{$z = 0.00338$} \\
\tableline
Ly$\alpha$  & 1215.670  & 1219.743 & \multicolumn{2}{c}{371$\pm$17} &  GHRS G160M pre-COSTAR data\\
\\
Ly$\beta$   & 1025.722  & 1029.158 & 170$\pm$16 & 158$\pm$17 & H$_2$ L6R3 contribution to listed W$_{obs}$ is $\approx25$ m\AA \\
\\
Ly$\gamma$  & \phn972.537 & \phn975.795 & \phn\phn0$\pm$33 & \phn33$\pm$14 \\
\\
\ion{O}{6}  & 1031.926    & 1035.383 & \phn26$\pm10$ & \phn25$\pm10$ & H$_2$ L6P4 nearby expected to be weak\\
	    & 1037.617    & 1041.093 & $<$30 & $<$30 & Blended with H$_2$ L5R3\\
\\
\ion{C}{3}  & \phn977.020 & \phn980.293 & 18$\pm$40 & \phn\phn9$\pm10$ \\
\cutinhead{$z = 0.00530$} 
Ly$\alpha$  & 1215.670  & 1222.150 & \multicolumn{2}{c}{367$\pm$13} & GHRS G160M pre-COSTAR data\\
Ly$\beta$   & 1025.722  & 1031.189 & 257$\pm$10 & 251$\pm$11 & H$_2$ L6P3 contribution to listed W$_{obs}$ is $\approx16$ m\AA \\
Ly$\gamma$  & \phn972.537 & \phn977.721 & 195$\pm25$ & 197$\pm$15 \\
Ly$\delta$  & \phn949.743 & \phn954.805 & 167$\pm$22 & 160$\pm$10 \\
Ly$\epsilon$& \phn937.804 & \phn942.802 & 133$\pm30$ & 135$\pm$10 & H$_2$ L15R3 nearby but does not interfere\\
Ly$\zeta$   & \phn930.748 & \phn935.709 & 121$\pm28$ & 127$\pm$10 & H$_2$ L16R3 contribution to listed W$_{obs}$ is $\approx11$ m\AA \\
Ly$\eta$    & \phn926.226 & \phn931.163 & \nodata & \nodata & In positive velocity wing of Galactic \ion{H}{1} Ly$\zeta$ \\
Ly$\theta$  & \phn923.150 & \phn928.070 & 110$\pm$26 & \phn85$\pm$13  \\
\\
\ion{O}{6}  & 1031.926    & 1037.426	& \nodata & \nodata & Blended with \ion{C}{2}$^*$ $\lambda1037.012$ and H$_2$ L5R1\\
	    & 1037.617    & 1043.147  & \phn17$\pm$10 & \phn\phn5$\pm$11\\
\\
\ion{C}{3}  & \phn977.020 & 982.227 & \nodata & $<74$ & Blended with H$_2$ L10R1\\
\\
\tableline
\\
\\
\\
\\
\cutinhead{$z = 0.02947$} 
Ly$\alpha$  & 1215.670  & 1251.496  & \multicolumn{2}{c}{139$\pm$07} & GHRS Ech-A post-COSTAR data\\  
	    & 		& 	    & \multicolumn{2}{c}{129$\pm$14} & GHRS G160M pre-COSTAR data \\
Ly$\beta$   & 1025.722  & 1055.950  & 29$\pm$10 & 15$\pm$10  \\
\\
\ion{O}{6}  & 1031.926   & 1062.337 & --5$\pm$10 & \phn5$\pm$10 \\
	    & 1037.617   & 1068.195 & \phn0$\pm$10 & \phn2$\pm$11 \\
\\
\ion{C}{3}  & \phn977.020 & 1005.813 & 10$\pm$10 & \phn3$\pm$10 & H$_2$ L8P2 nearby but does not interfere\\
\cutinhead{$z = 0.04898$} 
Ly$\alpha$  & 1215.670    & 1275.201 & \multicolumn{2}{c}{126$\pm$14} & GHRS G160M pre-COSTAR data\\
Ly$\beta$   & 1025.722  & 1075.952  & 23$\pm$10 & --2$\pm$22 & SiC2B \\
\\
\ion{O}{6}  & 1031.926    & 1082.459 & 46$\pm$43 & --6$\pm$26 & SiC1A, SiC2B\\
	    & 1037.617    & 1088.429 & --8$\pm$46 & --3$\pm$24 & SiC1A, SiC2B\\
\\
\ion{C}{3}  & \phn977.020 & 1024.865 & \nodata & \nodata & Line falls in wing of Galactic Ly$\beta$ \\
\cutinhead{$z = 0.06655$} 
Ly$\alpha$  & 1215.670    & 1296.573 & \multicolumn{2}{c}{312$\pm$13} & GHRS G160M pre-COSTAR data\\
Ly$\beta$   & 1025.722    & 1093.984 & \nodata & $125\pm10$ & Line falls in detector \#1 wavelength gap \\
	    &		  &	     & 	       & 	    & H$_2$ L1P1 contribution to listed W$_{obs}$ is $\approx25$ m\AA \\
	    &		  &	     & 	       & 	    & \ion{C}{3} at $z=0.12007$ is nearby \\
Ly$\gamma$  & \phn972.537 & 1037.259 & \nodata & \nodata & Blended with Galactic \ion{C}{2}/\ion{C}{2}$^*$/H$_2$ absorption\\
\\
\ion{O}{6}  & 1031.926    & 1100.601 & \phn3$\pm$10 & \phn6$\pm$10 \\
	    & 1037.617    & 1106.670 & \phn0$\pm$10 & --7$\pm$12 \\*
\\*
\ion{C}{3}  & \phn977.020 & 1042.041 & 10$\pm$10 & 21$\pm$15 \\*
\cutinhead{$z = 0.09012$} 
Ly$\alpha$  & 1215.670    & 1325.226 & \multicolumn{2}{c}{160$\pm$20} & GHRS G160M pre-COSTAR data, very broad line\\
Ly$\beta$   & 1025.722    & 1118.160 & 34$\pm$10 & 27$\pm$13 & Line very weak, subject to continuum placement  \\
\\
\ion{O}{6}  & 1031.926    & 1124.923 & $<$45 & $<$39 & Blended with \ion{Fe}{2} $\lambda$1133.665\\
	    & 1037.617    & 1131.127 & 17$\pm$10 & 12$\pm$10 & Subject to continuum placement \\
\\
\ion{C}{3}  & \phn977.020 & 1065.069 & $<$25 & $<$30 & No line present after removing H$_2$ L3R2 \\
\cutinhead{$z = 0.12007$} 
Ly$\alpha$  & 1215.670   & 1361.635  &  \multicolumn{2}{c}{138$\pm$11} & GHRS G160M pre-COSTAR data\\
Ly$\beta$   & 1025.722  & 1148.880   & 31$\pm$10 & 28$\pm$10\\
Ly$\gamma$  & \phn972.537 & 1089.309 & \phn0$\pm$42 & --3$\pm$13 & SiC1A, not covered by LiF1A\\
\\
\ion{O}{6}  & 1031.926    & 1155.829 & 33$\pm$11 & 27$\pm$10\\
	    & 1037.617    & 1162.204 & 18$\pm$10 & 17$\pm$10\\
\\
\ion{C}{3}  & \phn977.020 & 1094.331 & \nodata &  32$\pm$10 & Line falls in detector \#1 wavelength gap\\
            &             &          &         &            & H$_2$ L1R2 (ISM) and Ly$\beta$ ($z=0.06655)$ nearby \\
\cutinhead{$z = 0.14660$} 
Ly$\alpha$  & 1215.670    & 1393.887 & \multicolumn{2}{c}{355$\pm$20} & GHRS G160M pre-COSTAR data (deblended)\tablenotemark{d} \\
Ly$\beta$   & 1025.722  & 1176.093 & 67$\pm$13 & 78$\pm$16   \\
Ly$\gamma$  & \phn972.537 & 1115.111 & 15$\pm$10 & \phn5$\pm$11\\
\\
\ion{O}{6}  & 1031.926    & 1183.206 & 14$\pm$12 & \nodata & Outside $FUSE$ detector \#2 bandpass\\
	    & 1037.617    & 1189.732 & \nodata & \nodata & Outside $FUSE$ bandpass \\
\\
\ion{C}{3}  & \phn977.020 & 1120.251 & \phn2$\pm$13 & \phn5$\pm$11
\enddata
\tablenotetext{a}{Errors are $1\sigma$ estimates and limits are $3\sigma$ estimates
unless otherwise indicated. }
\tablenotetext{b}{Equivalent widths measured on $FUSE$ detectors 1 and 2.  Unless 
otherwise specified in the comments, values refer to LiF1/LiF2 for $\lambda > 990$\,\AA,
and SiC1/SiC2 for $\lambda < 990$\,\AA.}
\tablenotetext{c}{Notation for molecular hydrogen (H$_2$) lines is the same 
as specified for Table~2 - see note $a$ of that table.}
\tablenotetext{d}{Value for deblended profile shown in Figure~7.}
\end{deluxetable}

\clearpage
\newpage
\begin{deluxetable}{cccccccc}
\tablecolumns{8}
\tablewidth{0pt} 
\tablecaption{Intergalactic Medium Column Densities\tablenotemark{a}}
\tablehead{$z$ & $v_{helio}$ & $b$(\ion{H}{1}) & $\log N$(\ion{H}{1}) & $\log N$(\ion{O}{6}) & $\log N$(\ion{C}{3}) & $\frac{N(H I)}{N(O VI)}$ & Note\tablenotemark{b} \\
& (\kms) & (\kms) & (cm$^{-2}$) & (cm$^{-2}$) & (cm$^{-2}$) }
\startdata
0.00338 & \phn1015 & $30.3\pm^{4.3}_{3.8}$ & $14.41\pm^{0.10}_{0.10}$ & $13.32\pm^{0.13}_{0.21}$ & $<12.67$ & $12.3\pm5.7$ & 1,3\\
\\
0.00530& \phn1590 & $16.1\pm1.1$ & $15.85\pm^{0.10}_{0.08}$ & $<13.68$ & $<13.08$ & $>148$ & 2\\ 
\\
        &          & $17.5\pm1.5$ & $15.77\pm^{0.12}_{0.10}$ & \nodata & \nodata & $>123$ & 1\\
\\
0.02947 & \phn8840 & $23.8\pm^{22.2}_{9.4}$ & $13.55\pm^{0.17}_{0.12}$ & $<13.38$ & $<12.67$ & $>1.5$ & 1,4\\ 
\\
0.04898 & 14695    & $20.9\pm^{23.1}_{9.8}$ & $13.50\pm^{0.25}_{0.14}$ & $<13.79$ & \nodata & $>0.5$ & 1,4\\  
\\
0.06655 & 19965    & $25.9\pm^{3.2}_{2.5}$ &  $14.21\pm^{0.11}_{0.09}$ & $<13.38$ & $<12.71$ & $>6.8$ & 1\\
\\
0.09012 & 27035    & $21.2\pm^{...}_{7.4}$ & $13.64\pm^{0.14}_{0.14}$ & $<13.49$ & $<12.59$ & $>1.4$ & 1\\  
\\
0.12007 & 36020    & $15.5\pm^{16.3}_{4.2}$ & $13.61\pm^{0.16}_{0.15}$ & $13.38\pm^{0.18}_{0.12}$ & $12.45\pm^{0.19}_{0.36}$ & $1.7\pm^{0.6}_{0.8}$ & 1,5,6 \\  
\\
0.14660 & 43980    & $46.1\pm^{27.3}_{10.1}$ & $13.95\pm^{0.08}_{0.08}$ & $<13.46$ & $<12.71$ & $>3.1$ & 1\\  
\enddata
\tablenotetext{a}{Errors are 1$\sigma$ estimates unless specified otherwise.  Limits are 
3$\sigma$ estimates based upon the measured equivalent widths and the assumption of a 
linear curve of growth.}
\tablenotetext{b}{Notes:  (1) \ion{H}{1} curve-of-growth results include Ly$\alpha$.  
(2) \ion{H}{1} curve-of-growth results exclude Ly$\alpha$.  (3) \ion{O}{6} result based
on linear curve of growth.  Broad weak \ion{O}{6} $\lambda1031.926$
absorption is present in both LiF1A and LiF2B.  The $1\sigma$ error on N(\ion{O}{6}) was derived
from single channel values of the equivalent width.  Combination of data from both channels 
implies a significance of $3-4\sigma$.
(4) Upper limit on $b$(\ion{H}{1}) set by observed width of profile.
(5) \ion{O}{6} result based on curve of growth; best fitting $b$-value is 7.2 \kms, but higher 
values are allowed.  Errors span the column density range expected for b-values of 3--50 \kms.
(6) \ion{C}{3} column density limit assuming equivalent width limit of 30 m\AA. 
}
\end{deluxetable}

\clearpage
\newpage
\begin{deluxetable}{lccccc}
\tablecolumns{6}
\tablewidth{355pt} 
\tablecaption{Interstellar Medium Column Densities\tablenotemark{a}}
\tablehead{Species & $\log N$ & $\log N$ & $\log N$ & Method\tablenotemark{b} & Note\tablenotemark{c}\\
                   & (Negative) & (Positive) & (Total) }
\startdata
\ion{H}{1} & 20.15 & 19.45 & 20.23 & \nodata & 1 \\
	   & \nodata  & \nodata & $20.20\pm0.05$ & PF & 2\\
\\
H$_2$($J=0$) & \nodata	& \nodata &  $15.00\pm0.30$ & PF & 3\\
\\
H$_2$($J=1$) & \nodata	& \nodata &  $15.48\pm0.18$ & PF & 3 \\
\\
H$_2$($J=2$) & \nodata	& \nodata &  $14.76\pm0.12$ & COG, PF & 3\\
\\
H$_2$($J=3$) & \nodata	& \nodata &  $14.73\pm0.12$ & COG, PF & 3\\
\\
H$_2$($J=4$) & \nodata	& \nodata &  $<14.40$ & AOD\\
\\
\ion{O}{6} & \nodata & \nodata & $14.73\pm0.04$ & AOD & 4 \\
	   & 	     & 	       & $14.77\pm0.05$ & AOD & 5 \\
\\
\ion{Si}{2}  & $\sim14.90$ & $\sim14.68$ & $15.10\pm0.10$ & AOD, COG & 6\\
\\
\ion{P}{2} & $\sim13.55$ & $\sim12.84$ & $13.60\pm0.04$ & AOD \\
\\
\ion{P}{3} & $<13.70$ & $<13.40$ & $<13.88$ & AOD \\
\\
\ion{S}{2}  & $\sim15.33$  &  $\sim14.88$ & $15.46\pm0.06$ & AOD, COG & 7\\
\\
\ion{S}{3}  & \nodata & \nodata & $14.64\pm0.06$ & AOD \\
\\
\ion{S}{6} & \nodata & \nodata & $<13.36$ & AOD \\
\\
\ion{Ar}{1} & $\sim14.32$ & $\sim13.65$ & $14.40\pm^{0.19}_{0.13}$ & AOD, COG & 8\\
\\
\tableline
\\
\\
\\
\ion{Fe}{2} & $\sim14.89$ & $\sim14.40$ & $15.01\pm0.08$ & AOD, COG & 9\\
\\
\ion{Fe}{3} & $\sim14.23$ & $\sim13.77$ & $14.36\pm0.04$ & AOD\\
\enddata
\tablenotetext{a}{Errors are 1$\sigma$ estimates unless specified otherwise.  Limits are 
3$\sigma$ estimates based upon either the apparent optical depth limits or the 
measured equivalent widths and the assumption of a 
linear curve of growth.}
\tablenotetext{b}{(COG): Curve of growth fit to the measured
equivalent widths of the lines for each species.  The derived
Doppler parameters listed in the notes below are measures of the overall 
extent of profiles, not of the individual component widths.
(PF): Profile fitting for the H$_2$ lines.  The synthetic H$_2$ spectrum
was constucted as indicated in the text.
(AOD): Apparent optical depth method.  Column densities are direct integrations
of the apparent column density profiles constructed for each species.}
\tablenotetext{c}{Notes: \\ \noindent
(1) \ion{H}{1}: Column densities derived from integration of the NRAO
21\,cm profile measured by Murphy, Sembach, \& Lockman (unpublished).  
The emission was assumed to be optically thin.
\\ \noindent
(2) \ion{H}{1}: Column density derived from a profile fit to the radiation 
damping wings of the Ly$\beta$ line at 1025.7\,\AA.
\\ \noindent
(3) These H$_2$ values are appropriate for a Doppler parameter $b \approx
6\pm2$ \kms. The absorption is centered near $\sim$ +23 \kms.  
\\ \noindent
(4) \ion{O}{6}: Velocity range of integration is --160 \kms\ to +100 \kms.
\\ \noindent
(5) \ion{O}{6}: Velocity range of integration is --160 \kms\ to +240 \kms.
\\ \noindent
(6) \ion{Si}{2}: b$_{COG}$ = $32.2\pm2.5$ \kms.  Lines used include
$\lambda\lambda$1020.699, 1260.433 (HST), 1304.370 (HST), and 1526.707 (HST).
\\ \noindent
(7) \ion{S}{2}: b$_{COG}$ = $20.1\pm2.5$ \kms.  Lines used include
$\lambda\lambda$1250.584 (HST), 1253.811 (HST), 1259.519 (HST).
\\ \noindent
(8) \ion{Ar}{1}: b$_{COG}$ = $12.1\pm2.0$ \kms.  Lines used include
$\lambda\lambda$1048.220, 10660.660. 
\\ \noindent
(9) \ion{Fe}{2}: b$_{COG}$ = $28.0\pm2.5$ \kms.  Lines used include
$\lambda\lambda$1055.262, 1096.877, 1112.048, 1121.975, 1125.448, 1133.665, 
1143.226, 1144.938, 2586.650 (HST), and 2600.173 (HST).}
\end{deluxetable}

\clearpage
\newpage
\begin{deluxetable}{lc}
\tablecolumns{2}
\tablewidth{0pt}
\tablecaption{Ion Ratios and Elemental Abundances\tablenotemark{a}}
\tablehead{\multicolumn{2}{c}{Ratios}}
\startdata
N(\ion{P}{3})/N(P\,{\sc ii}+P\,{\sc iii}) 	& $<1.90$ 	 \\
\\
N(\ion{S}{3})/N(S\,{\sc ii}+S\,{\sc iii})	& $0.13\pm0.02$ \\
\\
N(\ion{Fe}{3})/N(Fe\,{\sc ii}+Fe\,{\sc iii})	& $0.18\pm0.04$	 \\
\cutinhead{Abundances\tablenotemark{b}}
(S/H) = N(\ion{S}{2})/N(\ion{H}{1})		   & $0.91\pm0.12$ (S/H)$_\odot$ \\
\\
(Ar/S) = N(Ar\,{\sc i})/N(S\,{\sc ii})	   & $0.45\pm0.13$ (Ar/S)$_\odot$\\
\\
(Si/S) = N(Si\,{\sc ii}+Si\,{\sc iii})/N(S\,{\sc ii}+S\,{\sc iii}) & $\ge0.20$ (Si/S)$_\odot$\\
\\
(P/S) = N(P\,{\sc ii}+P\,{\sc iii})/N(S\,{\sc ii}+S\,{\sc iii})  & $0.6-1.7$ (P/S)$_\odot$\\
\\
(Fe/S) = N(Fe\,{\sc ii}+Fe\,{\sc iii})/N(S\,{\sc ii}+S\,{\sc iii}) & $0.22\pm0.04$ (Fe/S)$_\odot$
\enddata
\tablenotetext{a}{Errors are $1\sigma$
estimates.  Limits are $3\sigma$ estimates.}
\tablenotetext{b}{Solar system reference abundances are from Anders \& Grevesse
(1989) on a logarithmic abundance scale where A(H) = 12.00, A(Si) = 7.55,
A(P) = 5.57, A(S) = 7.27, A(Ar) = 6.56, and A(Fe) = 7.51.  }
\end{deluxetable}

\clearpage
\newpage
\begin{deluxetable}{lcccccc}
\tablewidth{0pt}
\tablecaption{Galactic Halo Column Density Ratios}
\tablehead{Sight Line & $l(\degr)$ & $b(\degr)$ & 
$\frac{\mbox{N(O\,{\sc vi})}}{\mbox{N(N\,{\sc v})}}$ & 
$\frac{\mbox{N(O\,{\sc vi})}}{\mbox{N(C\,{\sc iv})}}$ & 
$\frac{\mbox{N(O\,{\sc vi})}}{\mbox{N(Si\,{\sc iv})}}$ & 
Notes\tablenotemark{a}}
\startdata
3C\,273      & 289.95 & $+64.36$   & $\phn7.2\pm1.1$ & $1.7\pm0.2$ & $8.9\pm1.1$ & 1 \\
\\
ESO\,141-55  & 338.18 & $-26.71$   & $\phn4.8\pm1.1$ & $0.6\pm0.2$ & $3.5\pm0.6$ & 2 \\
\\
PKS\,2155-304& \phn17.73&$-52.25$  & $\phn3.9\pm1.1$ & $1.6\pm0.2$ & \nodata     &   \\
\\
Mrk\,509     & \phn35.97&$-29.86$  & $\phn6.8\pm1.2$ & $1.8\pm0.2$ & $5.1\pm1.0$ & 3 \\
\\
H\,1821+643  & \phn94.00&$+27.42$  & $\phn2.6\pm0.5$ & $1.6\pm0.3$ & \nodata     & 4
\enddata
\tablenotetext{a}{All errors are $1\sigma$ estimates.  
Values of N(\ion{O}{6}) are from Savage et al. (2000), and 
values of N(\ion{C}{4}) and N(\ion{N}{5}) are from Savage et al. (1997) 
unless specified otherwise as follows:  
(1) Values are from this work and do not include \ion{O}{6} associated
with the positive velocity wing.  Including the positive velocity 
\ion{O}{6} wing increases the ratios by 10\%.  
(2) Sembach et al. (1999a).  A substantial saturation correction
was required for N(\ion{C}{4}); 
(3)  Values do not include the highly ionized high velocity clouds 
detected by Sembach et al. (1999b) and Sembach et al. (2000b).
N(\ion{Si}{4}) was measured from data obtained by Sembach et al. (1999b); 
(4) N(\ion{O}{6}) is from Oegerle et al. (2000).}
\end{deluxetable}


\end{document}